\documentclass[usenatbib]{mn2e}
\usepackage{graphicx}
\usepackage[space]{grffile}
\usepackage{latexsym}
\usepackage{amsfonts,amsmath,amssymb}
\usepackage{url}
\usepackage[utf8]{inputenc}
\usepackage{textcomp}
\usepackage{longtable}
\usepackage{multirow,booktabs}
\usepackage{natbib}
\usepackage[colorlinks, allcolors=blue]{hyperref}
\usepackage{pdflscape}
\usepackage[T1]{fontenc}
\usepackage{aecompl}
\usepackage{breakurl}
\usepackage{color}
\title[\emph{Herschel}-BAT Sample: SPIRE Catalog]{\emph{Herschel} Far-Infrared Photometry of the Swift Burst Alert
Telescope Active Galactic Nuclei Sample of the Local Universe. II. SPIRE
Observations\thanks{{\it Herschel} is an ESA space observatory with science instruments provided by European-led Principal Investigator consortia and with important participation from NASA.}}

\author[T. T. Shimizu]{T. Taro Shimizu$^{1}$\thanks{Email: tshimizu@astro.umd.edu}, Marcio Mel\'endez$^1$, Richard F. Mushotzky$^1$, Michael J. Koss$^{2}$, \newauthor Amy J. Barger$^{3,4,5}$, and Lennox L. Cowie$^{5}$\\
$^{1}$Department of Astronomy, University of Maryland, College Park, MD 20742, USA\\
$^{2}$Institute for Astronomy, Department of Physics, ETH Zurich, Wolfgang-Pauli-Strasse 27, CH-8093 Zurich, Switzerland\\
$^{3}$Department of Astronomy, University of Wisconsin-Madison, 475 N. Charter Street, Madison, WI 53706, USA\\
$^{4}$Department of Physics and Astronomy, University of Hawaii, 2505 Correa Road, Honolulu, HI 96822, USA\\
$^{5}$Institute for Astronomy, University of Hawaii, 2680 Woodlawn Drive, Honolulu, HI 96822, USA}

\newcommand{\herschel}{\textit{Herschel}}
\newcommand{\um}{$\mathrm{\mu}$m}
\newcommand{\swift}{\textit{Swift}}

\newcommand\aj{{AJ}}%
\newcommand\araa{{ARA\&A}}%
\newcommand\apj{{ApJ}}%
\newcommand\apjl{{ApJL}}%
\newcommand\apjs{{ApJS}}%
\newcommand\aap{{A\&A}}%
\newcommand\aaps{{A\&AS}}%
\newcommand\mnras{{MNRAS}}%
\newcommand\nar{{New A Rev.}}%
\newcommand\pasp{{PASP}}%
\newcommand\nat{{Nature}}%
\newcommand\ssr{\ref@jnl{Space~Sci.~Rev.}}%
  
\begin{document}



\maketitle

\label{firstpage}

\begin{abstract}
We present far-infrared (FIR) and submillimeter photometry from the \herschel{} \textit{Space Observatory's Spectral and Photometric Imaging Receiver} (SPIRE)  for 313 nearby ($z<0.05$) active galactic nuclei (AGN).  We selected AGN from the 58 month \swift{} Burst Alert Telescope (BAT) catalog, the result of an all-sky survey in the 14--195 keV energy band, allowing for a reduction in AGN selection effects due to obscuration and host galaxy contamination. We find 46\% (143/313) of our sample is detected at all three wavebands and combined with our PACS observations represents the most complete FIR spectral energy distributions of local, moderate luminosity AGN. We find no correlation between the 250, 350, and 500 \um{} luminosities with 14--195 keV luminosity, indicating the bulk of the FIR emission is not related to the AGN. However, Seyfert 1s do show a very weak correlation with X-ray luminosity compared to Seyfert 2s and we discuss possible explanations. We compare the SPIRE colors ($F_{250}/F_{350}$ and $F_{350}/F_{500}$) to a sample of normal star-forming galaxies, finding the two samples are statistically similar, especially after matching in stellar mass. But a color-color plot reveals a fraction of the \herschel-BAT AGN are displaced from the normal star-forming galaxies due to excess 500 \um{} emission ($E_{500}$). Our analysis shows $E_{500}$ is strongly correlated with the 14--195 keV luminosity and 3.4/4.6 \um{} flux ratio, evidence the excess is related to the AGN. We speculate these sources are experiencing millimeter excess emission originating in the corona of the accretion disk. 
\end{abstract}


\section{Introduction}\label{sec:intro}
The star formation rate (SFR) of galaxies sets the rate at which galaxies grow and evolve and is the one of the most important measures for understanding the hierarchical build-up of our universe over cosmic time. Large scale simulations, however, have shown that unregulated star formation leads to an overabundance of high mass galaxies \citep[e.g.][]{Bower:2006gf,Croton:2006kx,Silk:2012fj}. Therefore some process (or processes) must be able to stop, or ``quench,'' star formation before the galaxy grows to be too big.

The answer seems to lie in supermassive black holes (SMBH) which nearly all massive galaxies harbor in their centers. SMBHs grow through accretion of cold material (Active Galactic Nuclei; AGN), and the huge loss of gravitational energy of the cold material is converted into radiation that is evident across the whole electromagnetic spectrum and manifests itself as a bright point source in the nucleus of galaxies. The AGN can deposit this energy into the ISM of its host galaxy through jets \cite[e.g.][]{Fabian:2003ek,Best:2007vn,Lanz:2015bq} or powerful outflows that either heat the gas or remove it altogether, i.e. ``feedback''  processes \citep[e.g][]{Alatalo:2011lk,Veilleux:2013qq,Harrison:2014xe,Tombesi:2015fj}. 
  
Indirect evidence of this ``feedback'' has been observed through the simple, scaling relationships between the mass of the SMBH and different properties of the host galaxy such as the stellar velocity dispersion in the bulge, the bulge mass, and the bulge luminosity \citep[e.g.][]{Kormendy:1995mz,Ferrarese:2000gf,Marconi:2003ve,Haring:2004ly,Gultekin:2009ul,Kormendy:2013fj}. The relative tightness of these relationships suggests a strong coevolution of the host galaxy and SMBH. Much debate remains however as to the exact mechanism of AGN feedback and whether or not it plays a dominant role in the overall evolution of galaxies especially in light of new observations at both low and high $M_{\rm BH}$ that seem to deviate from the well-established relationships \citep[see][for a detailed review]{Kormendy:2013fj}. 

Evidence for AGN feedback though should also manifest itself in the SFR of its host galaxy, therefore much work has also focused on the so-called starburst-AGN connection \citep[e.g.][]{Sanders:1988fk,Cid-Fernandes:2001uq,Diamond-Stanic:2012rw,Dixon:2011yq,Rovilos:2012wd,Chen:2013uq,LaMassa:2013hb,Esquej:2014vl,Hickox:2014yq,Mushotzky:2014ad}. The problem lies in determining accurate estimates of the SFR in AGN host galaxies. Well-calibrated indicators, such as H$\alpha$ emission and UV luminosity, are significantly, if not completely, contaminated by the central AGN. Many studies therefore turn to the infrared (IR) regime ($1<\lambda<1000$ \um) where dust re-emits the stellar light from young stars.

Dust fills the interstellar medium (ISM) of galaxies and plays an important part in the heating and cooling of the ISM and the general physics of the galaxy. While dust contributes very little to the overall mass of a galaxy ($<1\%$), the radiative output, mainly in the infrared (IR) regime, can, on average, constitute roughly half of the bolometric luminosity of the entire galaxy \citep{Hauser_2001,Boselli_2003,Dale:2007fk,Burgarella_2013}, although there is an enormous range in the fraction. Dust efficiently absorbs optical and UV emission and re-radiates it in the mid- and far-infrared (MIR, FIR) depending on the temperature as well as grain size \citep{Draine:2003gd}. Recently formed O and B stars produce the majority of the optical and UV light in galaxies, therefore measuring the total IR light from dust provides insights into the current ($<100$ Myr) star formation rate (SFR) \citep[e.g.][]{Kennicutt:2012it}, although for very passive galaxies where the current SFR is much lower than it was earlier, IR emission can be an overestimate due to dust heating by an older stellar population.\citep[e.g.][]{Bendo:2015lr}

However, dust is also the key component in obscuring our view of AGN. Dust heated by the AGN is thought to primarily live in a toroidal-like structure that encircles the AGN and absorbs its radiative output for certain lines of sight. The dusty torus is used to explain the dichotomy of AGN into Seyfert 1 (Sy 1) and Seyfert 2 (Sy 2) within a unified model \citep{Antonucci:1993os,Urry:1995il}. Like O and B stars in star-forming regions, the AGN outputs heavy amounts of optical and UV light, and like dust in the ISM the dusty torus absorbs and re-emits this as IR radiation. Spectral energy distribution (SED) models \citep{Barvainis:1987ty,Pier:1992sf,Efstathiou:1995rz,Nenkova:2002ys,Fritz:2006yq}  as well as observations \citep{Elvis:1994uq,Spinoglio:2002uq,Netzer:2007ve,Mullaney:2011yq,Mor:2012fj} suggest the torus mainly emits in the MIR ($3<\lambda<40$ \um) with the flux density dropping rapidly in the FIR ($\lambda>40$ \um). Further the SED for stellar dust re-radiation peaks in the FIR \citep{Calzetti:2000fk,Dale:2002ty,Draine:2007rm}, making the FIR the ideal waveband to study star-formation in AGN host galaxies.

Space-based telescopes such as the \textit{Infrared Astronomical Satellite} \cite[IRAS;][]{Neugebauer:1984fp}, \textit{Spitzer Space Telescope} \citep{Werner:2004cr}, and \textit{Infrared Space Observatory} \citep{Kessler:1996wd} greatly expanded our knowledge of the IR universe and provided a window into the FIR properties of galaxies. But, before the launch of the \textit{Herschel Space Observatory} \citep{Pilbratt:2010rz}, the FIR SED was limited to $\lambda < 200$ \um, except for studies of small samples of the brightest galaxies using ground-based instruments such as \textit{SCUBA} \citep[e.g.][]{Papadopoulos:1999lr, Papadopoulos:2000fk}. \herschel{} with the Spectral and Photometric Imaging Receiver \citep[SPIRE;][]{Griffin:2010sf} has pushed into the submillimeter range with observations in the 250, 350, and 500 \um{} wavebands, probing the Rayleigh-Jeans tail of the modified blackbody that accurately describes the broadband FIR SED of galaxies \citep[e.g.][]{Calzetti:2000fk,Dale:2012dq,Cortese:2014qq}. These wavebands are crucial for measuring dust properties (i.e. temperature and mass) as \citet{Galametz:2011ao} and \citet{Gordon:2010ix} show. Further, \citet{Ciesla:2015qr} found that FIR and submillimeter data are important for estimating the SFR of AGN host galaxies. 

Recent studies, such as \citet{Petric:2015fk} and  \citet{Xu:2015yq}, focusing on the dust and star-forming properties of AGN have shown the power of long wavelength \herschel{} data to better constrain the SFR, dust mass, and dust temperature in AGN host galaxies. \citet{Petric:2015fk} analyzed the IR SEDs of low redshift ($z<0.5$), quasi-stellar objects (QSOs) broadly finding most of the FIR emission can be attributed to thermally-heated dust. \citet{Xu:2015yq} looked at the IR SEDs of 24 $\mu$m selected AGN at slightly higher redshift ($0.3<z<2.5$) around galaxy clusters finding a strong correlation between the AGN and star-forming luminosity which could be due to their shared correlation with galaxy stellar mass. Both studies, however, rely on AGN selection using different wavebands (optical vs. mid-infrared) and generally probe the higher AGN luminosity population.

Therefore, we have assembled a large ($\sim300$), low redshift ($z<0.05$) sample of AGN selected using ultra-hard X-ray observations with the \swift/\textit{Burst Alert Telescope} (BAT) and imaged each one with \herschel. This sample focuses on moderate luminosity Seyfert galaxies ($10^{42} < L_{\rm AGN} < 10^{46}$).  In \citet{Melendez:2014yu}, we presented the PACS data of the \herschel-BAT AGN which provided photometry at 70 and 160 \um. In this paper, we complete the FIR SED of the BAT AGN with the creation and analysis of the SPIRE images. We focus on the overall luminosity distributions at the SPIRE wavebands as well as the SPIRE colors ($F_{250}/F_{350}$ and $F_{350}/F_{500}$) to determine the likely heating sources of cold dust in AGN host galaxies. We also look for correlations with a proxy for the bolometric AGN luminosity to potentially reveal any indication that AGN heated dust is contributing to the FIR SED. This paper sets us up for a complete study of the mid-far IR SED to fully explore the star-forming properties of AGN host galaxies and reveal the global starburst-AGN connection in the nearby universe (Shimizu et al, in preparation). Throughout this paper we assume a $\Lambda$CDM cosmology with $H_0=70$ km s$^{-1}$ Mpc$^{2}$, $\Omega_{m} = 0.3$, and $\Omega_{\Lambda}=0.7$.  Luminosity distances for each AGN were calculated based on their redshift and assumed cosmology, except for those with $z < 0.01$ where we referred to the \textit{Extragalactic Distance Database}\footnote{\url{http://edd.ifa.hawaii.edu/}}.
  
\section{The \swift/BAT AGN Sample}\label{sample}
\swift/BAT \citep{Barthelmy_2005,Gehrels_2004} operates in the 14--195 keV energy range, continuously monitoring the sky for gamma-ray bursts. This constant monitoring has also allowed for the most complete all-sky survey in the ultra-hard X-rays. To date, BAT has detected 1171 sources at $>4.8\sigma$ significance corresponding to a sensitivity of $1.34\times10^{-11}$ ergs s$^{-1}$ cm$^{-2}$ \citep{Baumgartner:2013fq}. Over 700 of those sources have been identified as a type of AGN (Seyfert, Blazar, QSO, etc.)

We selected our sample of 313 AGN from the 58 month \swift/BAT Catalog\footnote{\url{https://swift.gsfc.nasa.gov/results/bs58mon}} \citep{Baumgartner:2012gf}, imposing a redshift cutoff of $z<0.05$. All different types of AGN were chosen only excluding Blazars/BL Lac objects which most likely introduce complicated beaming effects. To determine their AGN type, for 252 sources we used the classifications from the BAT AGN Spectroscopic Survey (Koss et al, in preparation) which compiled and analyzed optical spectra for the \swift/BAT 70 month catalog \citep{Berney:2015lr}. Seyfert classification was determined using the standard scheme from \citet{Osterbrock:1977fj} and \citet{Osterbrock:1981uq}. For the remaining 61 AGN we used the classifications provided in the 70 month catalog. In total the sample contains 30 Sy 1, 30 Sy 1.2, 79 Sy 1.5, 1 Sy 1.8, 47 Sy 1.9, 121 Sy 2, 4 LINERs, and 1 unclassified AGN. For the purpose of broad classification, in the rest of this paper we choose to classify all Sy 1-1.5 as Sy 1's, and all Sy 1.8-2 as St 2's. In Table~\ref{tbl:bat_info} we list the entire \herschel{}-BAT sample along with positions and redshifts taken from the \textit{NASA/IPAC Extragalactic Database} (NED)\footnote{\url{http://ned.ipac.caltech.edu/}}.

Selection of AGN by ultra-hard X-rays provides multiple advantages over other wavelengths. Due to their high energy, ultra-hard X-rays easily pass through Compton-thin gas or dust in the line of sight providing a direct view of the AGN. Using optical or mid-infrared selection can be problematic due to contamination by the host galaxy. Also, ultra-hard X-rays are unaffected by any type of absorption by material obscuring the AGN provided it is optically thin to Compton scattering ($N_{\rm{H}} \lesssim 10^{24}\,\, \rm{cm}^{-2}$) which is a concern for hard X-rays in the 2-10 keV energy range.

Numerous studies have been done on the BAT sample in the past that span nearly the entire electromagnetic spectrum. \citet{Weaver:2010rt} and \citet{Melendez:2008pd} used \textit{Spitzer}/IRS spectra to study the mid-infrared properties of the BAT AGN. \citet{Winter:2009kx} and~\citet{Vasudevan:2013dz} studied the X-ray spectral properties for a subsample, while \citet{Koss:2011vn} looked at the optical host galaxy properties and \citet{Winter:2010yq} analyzed the optical spectra. Along with these, many of the BAT AGN are detected at radio wavelengths with the FIRST \citep{Becker:1995lq} and NVSS \citep{Condon:1998eu} survey as well. One key ingredient missing though is the far-infrared (FIR) where emission from ultraviolet-heated dust peaks.
%
\begin{table*}
\begin{minipage}{6.0in}
\caption{The \herschel-BAT Sample}
\label{tbl:bat_info}
\begin{tabular}{@{}lllcclcc}
\hline
Name & RA & DEC & $z$ & Distance &Type  & OD & OBSID\\
           & (J2000) & (J2000) &  & (Mpc) &  &  & \\
\hline
Mrk 335	&	00h06m19.5s	&	+20d12m10s	&	0.0258	&	112.62	&	Sy 1.2	&	949	&	1342234683	\\
2MASX J00253292+6821442	&	00h25m32.9s	&	+68d21m44s	&	0.012	&	51.87	&	Sy 2	&	1022	&	1342239794	\\
CGCG 535-012	&	00h36m21.0s	&	+45d39m54s	&	0.0476	&	211.45	&	Sy 1.2$^{a}$	&	976	&	1342237509	\\
NGC 235A	&	00h42m52.8s	&	-23d32m28s	&	0.0222	&	96.83	&	Sy 2	&	737	&	1342221462	\\
MCG -02-02-095	&	00h43m08.8s	&	-11d36m04s	&	0.0189	&	81.99	&	Sy 2	&	949	&	1342234695	\\
Mrk 348	&	00h48m47.1s	&	+31d57m25s	&	0.015	&	65.13	&	Sy 1.9	&	603	&	1342212368	\\
MCG +05-03-013	&	00h51m35.0s	&	+29d24m05s	&	0.036	&	158.22	&	Sy 1$^{a}$	&	977	&	1342237559	\\
Mrk 352	&	00h59m53.3s	&	+31d49m37s	&	0.0149	&	64.39	&	Sy 1.2	&	964	&	1342236244	\\
ESO 195-IG021 NED03	&	01h00m35.0s	&	-47d52m04s	&	0.0482	&	214.22	&	Sy 2	&	949	&	1342234727	\\
MCG -07-03-007	&	01h05m26.8s	&	-42d12m58s	&	0.0299	&	130.92	&	Sy 2	&	949	&	1342234725	\\
2MASX J01064523+0638015	&	01h06m45.3s	&	+06d38m02s	&	0.041	&	181.09	&	Sy 2	&	976	&	1342237549	\\
\hline
\end{tabular}
\medskip
Note. -- \textit{Column 1:} Name of the source. \textit{Column 2:} Right ascension in J2000 coordinates. \textit{Column 3:} Declination in J2000 coordinates. \textit{Column 4:} Redshift of the source. \textit{Column 5:} Luminosity distance of the source in Mpc. \textit{Column 6:} AGN classification. Sources marked with an $^{a}$ are from the 70 month \textit{Swift}/BAT catalog; all others are from the BASS survey \citep{Berney:2015lr}. \textit{Column 7:} \textit{Herschel} Operational Day number for when the observation started. \textit{Column 7:} \textit{Herschel} Observation Identification number. The full table is available in the online version.
\end{minipage}
\end{table*}
%
\section{\herschel{} SPIRE Observations and Data Reduction}\label{sec:obs}
The Spectral and Photometric Imaging Receiver (SPIRE) \citep{Griffin:2010sf} onboard \herschel{} observed in small map mode 293 of our objects between operational days (OD) 722 and 1265 as part of a Cycle 1 open time program (OT1\_rmushotz\_1, PI: Richard Mushotzky). 20 other objects with public data from separate programs are also included to complete our sample. Within each observation from our program, two scans were performed at nearly orthogonal angles with the nominal 30" s$^{-1}$ scan speed that resulted in a $\sim$5' diameter area of homogeneous coverage in all three SPIRE wavebands centered at 250, 350, and 500 \um. Table 1 lists the OD and OBSID for each source.

The SPIRE raw data (``Level 0'') were reduced to ``Level 1'' using the standard pipeline contained in the \textit{Herschel Interactive Processing Environment} (HIPE) version 13.0 \citep{Ott:2010rm}. The pipeline performs a host of steps including, but not limited to, glitch removal, electrical crosstalk correction, and brightness conversion, which results in timeline data (brightness vs. time) for each bolometer and each scan. 

\begin{figure*}
\begin{center}
\includegraphics[width=0.8\textwidth]{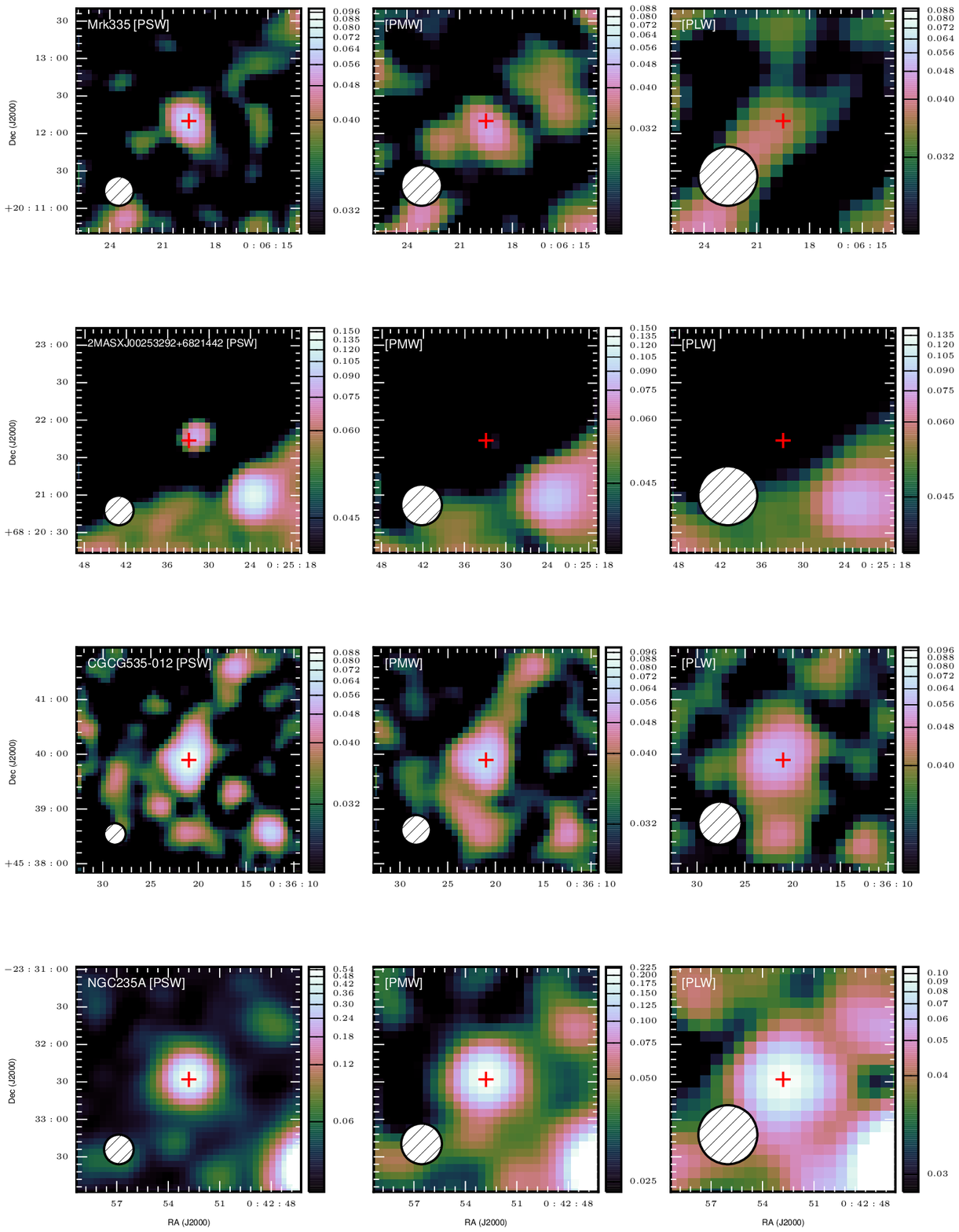}
\caption{\label{fig:example_maps} SPIRE maps for all of the \herschel-BAT AGN (Here we only show the first four in the sample. Figures for all of the sources are available online.) Each row is a separate AGN and each column is a separate waveband. Left: PSW or 250 \um. Middle: PMW or 350 \um. Right: PLW or 500 \um. All images have been smoothed for aesthetic purposes using a 2D Gaussian kernel with $\sigma=1.5$ pixels, slightly smaller than the beam size. Pixel intensity units are Jy/beam and are displayed with an arcsinh stretch. The range in intensity is from the maximum pixel value near the BAT position or 5 times the median global background (whichever is larger) to the median global background level measured during the aperture photometry process. The white hatched circle in the lower left corner shows the FWHM of the \herschel{} beam convolved with the same Gaussian as the image. The red cross plots the known position of the BAT AGN from Table~\ref{tbl:bat_info}.}
\end{center}
\end{figure*}

The Level 1 timelines were then input into \textit{Scanamorphos} v24.0 \citep{Roussel:2013gf} to create image maps for each source. \textit{Scanamorphos} was effectively designed to take advantage of the built-in redundancy of the detectors to subtract the low frequency noise caused by temperature drifts of the telescope as a whole (correlated noise) and each bolometer. The drifts are determined from the data themselves without the use of any noise model and thus more accurately take into account any time variation of the drifts. The final output of \textit{Scanamorphos} is a FITS image cube or series of FITS files containing the image, 1$\sigma$ pixel error, drifts, weights, and clean map. Each map has pixel sizes equal to $ \sim$1/4 times the point spread function (PSF) full width at half maximum (FWHM) of each waveband. For the 250 (18" FWHM), 350 (24" FWHM), and 500 \um{} (36" FWHM) maps, this means 4.5", 6.25", and 9" pixel sizes respectively. The brightness units for the maps are Jy/beam. Fig.~\ref{fig:example_maps} shows the resulting maps, centered on the known positions of the AGN from Table~\ref{tbl:bat_info}.

\section{SPIRE Flux Extraction}\label{photom}
Because of the large beams of SPIRE, a subsample of sources remain unresolved even though all are fairly nearby ($\rm{z} < 0.05$).  We employed two different methods depending on whether a source is resolved (aperture photometry) or unresolved (timeline fitting). 

\subsection{Timeline Fitting}\label{timeline}
Timeline fitting involves modeling the response of a point source in the Level 1 data as a Gaussian and determining the best fit parameters for the Gaussian. The peak of the Gaussian then corresponds to the flux density of the source. Because this method is performed on the Level 1 data, instead of the image maps, it avoids any potential artifacts or biases involved with the mapmaking procedure and is the highly recommended procedure for determining the photometry of point sources by the SPIRE Data Reduction Guide (DRG, section 5.7.1)\footnote{\url{http://herschel.esac.esa.int/hcss-doc-11.0/index.jsp\#spire_drg:_start}}.

To determine which sources are unresolved, we fit the Level 1 data using the \texttt{sourceExtractorTimeline} task within HIPE to measure the best-fit Gaussian where one of the free parameters is the size of the source, represented by the FWHM of the Gaussian. A source is then considered unresolved in a waveband if its best fit FWHM is less than 21", 28", or 40" at 250, 350, or 500 \um{} respectively, the upper limit for the nominal ranges of FWHM expected for point sources. We also visually inspected the images to ensure no extended sources were falsely classified as a point source. This occurred when an extended galaxy contained a bright point source in the nucleus. To avoid combining different flux extraction techniques for a single source, we only used the timeline flux densities if that source was unresolved at all wavebands in which it is visually detected.

We used the timeline fluxes for 82 (26\%), 62 (20\%), and 12 (13\%) sources in each of the three wavebands. These sources are indicated in Table~\ref{tbl:spire_flux} with a ``TF". The discrepancy is due to some of the sources being undetected at longer wavelengths because of the rapid fall-off of the SED as well as the decreasing sensitivity of SPIRE. For the sources that are undetected we used aperture photometry to determine their $5\sigma$ upper limit, therefore if a source is detected as a point source at 250 \um{} but undetected at 350 and 500 \um, it would be listed as having a timeline flux for 250, but not for 350 and 500. For reference all but three of the 82 sources for which we used timeline fitting to determine the 250 \um{} photometry are also point-like in both PACS wavebands. The exceptions, however are only partially resolved at 70 \um{} and point-like at 160 \um{}. 

\subsection{Aperture Photometry}\label{aperture}
For the rest of the sources, we perform aperture photometry to measure the flux densities directly from the \textit{Scanamorphos}-produced SPIRE maps. The first step in aperture photometry is to determine the size and shape of the aperture from which to extract the flux from. To determine the local background, we also used a concentric annulus around the source aperture as well as a series of apertures within the annulus to calculate the background root-mean-square (RMS).

Instead of choosing apertures manually by visually inspecting each image, we used the publicly available, Python based \texttt{photutils}\footnote{\url{http://photutils.readthedocs.org/en/stable/}} package. \texttt{photutils} provides open-source functions that perform tasks such as detecting sources, measuring their size and shape, and performing aperture photometry. The process we used for the aperture photometry of the BAT AGN in the SPIRE maps involved the following key steps and is very similar to the method employed in the popular \texttt{Sextractor} software \citep{Bertin:1996fk} especially in regards to the use of a segmentation image.

\begin{enumerate}
    \item Convert the maps from Jy/beam to Jy/pixel.
    \item Measure the standard deviation and median of the global background level.
    \item Detect sources above a given threshold using a segmentation image--details are given in the next section.
    \item Find the associated BAT source.
    \item Measure the size and shape of associated source.
    \item Create the source and background annulus from the size and shape of the source.
    \item Create a series of background apertures around the source aperture to measure the RMS of the background.
    \item Measure the fluxes within all apertures and calculate a background-subtracted flux and uncertainty.
\end{enumerate}

\subsubsection{Source Detection}
The first step in the process is converting the SPIRE map units from Jy/beam to Jy/pixel. The images must be divided by the beam area specific to the waveband and calibration version (spire\_cal\_13\_1) used to create the maps. For this work the beam areas are 469.7, 831.7, and 1793.5 arcsec$^{2}$ for the 250, 350, and 500 \um{} wavebands respectively and taken from the latest version of the SPIRE DRG. Each pixel of the map is converted to Jy/pixel using the following formula.

\begin{equation}
I\,[Jy/pixel] = \frac{I\,[Jy/beam] \times P^2}{B}
\end{equation}

\noindent $I$ is the intensity value of the pixel, $P$ is the pixel size of the map (see Section~\ref{sec:obs}), and $B$ is the beam area stated above. After converting all of the pixels, we use an iterative procedure to measure the median and standard deviation of the background over the whole map. For this, we use two tools: sigma-clipping and a segmentation image. Sigma-clipping involves measuring the median and standard deviation of data (in this case pixel values of the SPIRE maps) and removing pixels that are above a clipping limit. The process is then repeated until there are no more pixels above the clipping limit. We chose a clipping limit of 3 standard deviations above the measured median. The function used to perform the sigma-clipping is \texttt{sigma\_clipped\_stats} that is provided within the \texttt{Astropy} \citep{Astropy:2013ek} package. 

Sigma-clipping however can still be affected by sources in the field and provides a biased estimate of the background. A better process is to iteratively run sigma-clipping, each time masking out pixels associated with a source. To determine which pixels will be masked, we use a segmentation image. A segmentation image is a map, the same size as the input map, that identifies groups of connected pixels that are above a certain threshold. For a threshold we use $MD + 2\times SD$ where $MD$ and $SD$ are the median and standard deviation of the map determined through sigma-clipping. A source is identified in the map as a group of 5 interconnected pixels that are above this threshold value. The \texttt{Photutils} function \texttt{detect\_sources} was used to create all segmentation images. All of the pixels that are associated with a source are then masked out and sigma-clipping is re-run on the remaining pixels. This process is repeated until the percentage change in the sigma-clipped median is less than $1\times10^{-6}$ or a maximum of 10 iterations. A final sigma-clipped median ($MD_{final}$) and standard deviation ($SD_{final}$) is measured from the masked map.

We then produce a new segmentation image to find the associated BAT source in the SPIRE map using a threshold of $MD_{final} + 1.5\times SD_{final}$. Through tests of various extended sources, we found $1.5\times SD_{final}$ to best incorporate the fainter outer regions of the galaxies. The \texttt{Photutils} function \texttt{segment\_properties} is then used to measure centroid, semimajor axis length, semiminor axis length, and position angle of all sources detected from the segmentation image. We identify the BAT source as the closest detected source within one FWHM (see Section~\ref{sec:obs}) of the known positions (Table~\ref{tbl:bat_info}). 

\subsubsection{Target and Background Apertures}
After the SPIRE source that is associated with the BAT source is found, we used the measured size and shape from \texttt{segment\_properties} to construct a target and background aperture. The target aperture is an ellipse and the background aperture is an elliptical annulus. The semimajor and semiminor radii of the target aperture are calculated as 3.5 times the semimajor and semiminor sigma values from \texttt{segment\_properties}. The sigma values are measured from the second-order central moments of the detected source and represent the standard deviations along each axis of a 2D Gaussian that has the same second-order moments. The central position of the target aperture is the centroid of the source and the orientation is the same as the measured orientation.

The background annulus has the same central position and orientation as the target aperture. For the inner radius, we increase the semimajor and semiminor axis of the target aperture by 3 pixels. The outer radius is then 1.5 times the inner radius.

\begin{figure*}
\begin{center}
\includegraphics[width=\textwidth]{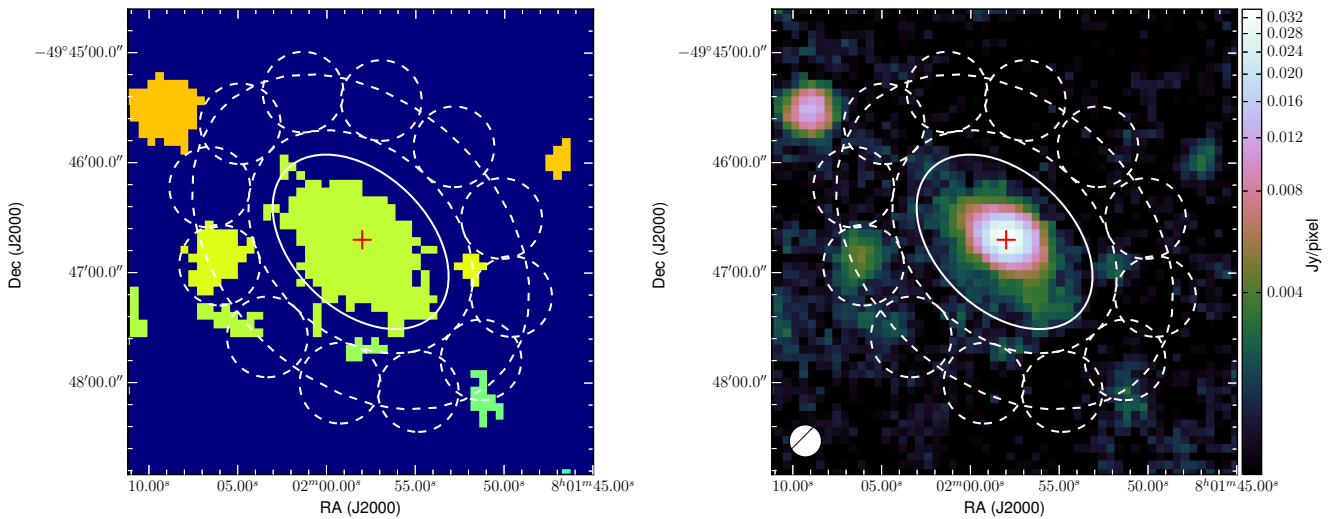}
\caption{\label{fig:example_photometry} An example of the segmentation image (left) and apertures used in the aperture photometry procedure. The solid white line indicates the target aperture whose size and shape were determined from the centroid and second order moments of the source (central light green group of pixels)  detected in the segmentation image. The dashed annulus shows the background annulus used to calculate the local background level. The dashed circular apertures are used to determine the background noise. The right image is the actual SPIRE map of the BAT source ESO 209-G012 using an arcsinh stretch. The red cross in both images shows the known location of ESO 209-G012 from Table~\ref{tbl:bat_info}.}
\end{center}
\end{figure*}

In addition to the background annulus, we also construct a series of circular apertures that encircle the target aperture. These have a size of 22", 30", and 42", the recommended size of an aperture (see the SPIRE DRG) for measuring the flux of a point source in the 250, 350, and 500 \um{} maps respectively. While the background annulus is used to measured the local background level, these circular apertures are used to measure the background noise. The SPIRE DRG recommends using local background apertures for the calculation of the background noise because calculating the RMS within the background annulus will underestimate the noise. We construct as many apertures as can fit just outside the target aperture without overlapping but impose a minimum of 6 apertures. Figure~\ref{fig:example_photometry} shows an example of the apertures used in calculating the photometry as well as the segmentation image that was used to find the source and determine its properties to construct the apertures. 

One exception to all of this occurs for small sources. If the constructed target aperture has a semimajor axis smaller than 22", 30", or 42" for 250, 350, and 500 \um{} maps, then we use a circular aperture with these radii. This indicates the source is likely a point source that was either missed using the results from the timeline fitting (Section~\ref{timeline}) or is extended at other SPIRE wavelengths which automatically identifies it as extended at all wavelengths. For these aperture photometry point sources, the background annulus used has a 60" inner radius and a 90" outer radius, the recommended annulus for point source photometry from the SPIRE DRG. The circular background apertures are still constructed in the same way as for extended sources.

The other exception is for sources that lie in maps dominated by foreground cirrus emission. Cirrus emission comes from cold dust in the Milky Way galaxy that is along our line of sight to the AGN. It is identified as bright smooth patches that occur over large spatial scales. We visually identified 25 sources that are likely contaminated by Milky Way cirrus. We used point source apertures for the photometry, however instead of placing the background annulus 60--90" away, we placed it right outside the target aperture to get a more accurate estimate of the local background. 

\subsubsection{Flux Extraction}
We calculate the raw source flux ($F_{raw}$) by summing the values of the pixels within the target aperture. Pixels that are on the border are used by determining the fraction of the pixel area that is inside the aperture and using that fraction of the pixel value in the sum. The background level is determined by estimating the mode of the pixel values within the background annulus using a Python version of the ``MMM'' routine which is the method used in the popular photometry package \texttt{DAOPHOT}. The mode ($F_{bkg-an}$) then represents the per-pixel background level so we multiply it by the pixel area of the target aperture ($A_{src}$) to calculate the total background flux within the target aperture. The background flux is then subtracted from the raw source flux. The whole procedure can be represented with the following equation:

\begin{equation}
F_{\rm{bkg-sub}} = F_{\rm{raw}} - F_{bkg-an} \times A_{src}
\end{equation}

For extended sources, $F_{\rm{bkg-sub}}$ represents the final measured flux density. However for sources which used the point source aperture, we applied the necessary aperture corrections as given in the SPIRE DRG. For the 250, 350, and 500 \um{} bands, these corrections are 1.2697, 1.2271, and 1.2194 respectively. 

Both \citet{Ciesla:2012lq} and \citet{Dale:2012dq} found aperture corrections for extended emission to be small and unnecessary.  To confirm this, we convolved our PACS 160 \um{} images (PSF FWHM 12") to the 250, 350, and 500 \um{} angular resolution using the convolution kernels from \citet{Aniano:2011rr}. This makes the assumption that the 160 \um{} emission is generated by the same material as that producing the SPIRE emission. Aperture corrections were calculated by dividing the total flux within an aperture from the original PACS image by the flux within the exact same aperture applied to the convolved image. The same aperture sizes were used in this calculation as the ones used in this SPIRE analysis. Median aperture corrections of 1.01, 0.98, and 0.98 were found consistent with a value of 1 and confirming that extended emission aperture corrections are not necessary. 

\subsection{Uncertainty Calculation}
For sources where we used aperture photometry, three components were factored into the total error budget for the SPIRE aperture photometry of our sample. These were the instrumental error ($err_{\rm{inst}}$), background error ($err_{\rm{bkg}}$), and calibration error ($err_{\rm{cal}}$). $err_{\rm{cal}}$ is fixed at 6.5\% of the measured background-subtracted flux density for sources which used aperture photometry. The calibration error is the combination of the 4\% uncertainty in the Neptune (which is the primary calibrator source for SPIRE) flux model, the 1.5\% uncertainty from repeated measurements of Neptune, and the 1\% uncertainty in the beam areas \citep{Bendo:2013sd}. To determine $err_{\rm{inst}}$, we summed in quadrature all of the 1$\sigma$ pixel uncertainties from the error map contained in the target aperture. For $err_{\rm{bkg}}$, we measured the flux within the circular background apertures placed around the source aperture. The standard deviation of the fluxes was calculated after using sigma-clipping with a 3$\sigma$ cutoff to remove fluxes possibly contaminated with a bright, background source.  This was then scaled to the area of the target aperture to represent $err_{\rm{bkg}}$.  The three error components are then summed in quadrature to form the total 1$\sigma$ uncertainty ($err_{\rm{tot}}$) of the measured flux density for each source.

For sources where we used the timeline fitting, only two components are needed. The output from the timeline fitting contains an estimate of the statistical uncertainty in the flux density. This is combined in quadrature with a 5.5\% calibration error, which is the same as the calibration error for aperture photometry minus the 1\% uncertainty in the beam areas that are not needed in the timeline fitting.

\section{The \herschel-BAT SPIRE Catalog}
Table~\ref{tbl:spire_flux} represents our final SPIRE catalog for the \herschel-BAT AGN. For each waveband three columns are provided. The first column contains the flux density ($F_{bkg-sub}$ and $1\sigma$ uncertainty ($err_{tot}$). The second column provides the photometry method used to determine the flux density, either timeline fitting (``TF'') point source aperture photometry (``PAP''), or extended source aperture photometry (``EAP''). The third column provides flags to assist in assessing the reliability of the photometry. We decided to impose a strict 5$\sigma$ threshold for reporting the photometry, so for all sources where $5 err_{\rm{tot}} > F_{\rm{bkg-sub}}$, only the 5$\sigma$ upper limit is given as the flux density for that band and a flag of ``U'' is used. For sources above $5\sigma$ a flag of ``A'' is used. Alongside these two flags we also indicate those sources that are contaminated by foreground cirrus emission with a flag of ``C''. Finally a flag of ``d'' or ``D'' is used for sources that have a nearby companion that could possibly be affecting the photometry of the main BAT source. ``d'' represents companions that are either relatively faint compared to the BAT source or are far enough away where contamination to the SPIRE photometry is minimal. ``D'' represents nearby bright companions that are completely contaminating the source photometry, and we recommend using these flux densities as only upper limits. In total 17 and 10 sources have a ``d'' and ``D'' classification for the 250 \um{} waveband, 20 and 13 for the 350 \um{} waveband, and 13 and 21 for the 500 \um{} waveband. The changing numbers with wavelength represents the degrading resolution as wavelength increases. Many of the sources with a ``d'' or ``D'' flag have previously been identified as merging or companion systems in \citet{Koss:2010nr}.

\begin{landscape}
\begin{table}
\centering
\begin{minipage}{9.6in}
\caption{SPIRE Flux Densities}
\label{tbl:spire_flux}
\begin{tabular}{@{}lcccllcllcll}
\hline
Name & RA & DEC & $F_{250}$ & Method$_{250}$ & Flag$_{250}$  & $F_{350}$ & Method$_{350}$ & Flag$_{350}$ &$F_{500}$ & Method$_{500}$ &Flag$_{500}$\\
 & (J2000) & (J2000) & (Jy) &  &  & (Jy) &  &  & (Jy) &  & \\
 \hline
Mrk 335	&	00h06m19.5s	&	+20d12m10s	&	$	0.068	\pm	0.007	$	&	TF	&	A	&	$		<	0.067	$	&	PAP	&	U	&	$		<	0.078	$	&	PAP	&	U	\\
2MASX J00253292+6821442	&	00h25m32.9s	&	+68d21m44s	&	$		<	0.286	$	&	PAP	&	UC	&	$		<	0.364	$	&	PAP	&	UC	&	$		<	0.263	$	&	PAP	&	UC	\\
CGCG 535-012	&	00h36m21.0s	&	+45d39m54s	&	$	0.178	\pm	0.034	$	&	EAP	&	A	&	$		<	0.140	$	&	EAP	&	U	&	$	0.061	\pm	0.011	$	&	EAP	&	A	\\
NGC 235A	&	00h42m52.8s	&	-23d32m28s	&	$	0.955	\pm	0.081	$	&	EAP	&	A	&	$	0.399	\pm	0.060	$	&	EAP	&	A	&	$		<	0.126	$	&	PAP	&	U	\\
MCG -02-02-095	&	00h43m08.8s	&	-11d36m04s	&	$		<	0.102	$	&	EAP	&	U	&	$		<	0.057	$	&	EAP	&	U	&	$		<	0.094	$	&	PAP	&	U	\\
Mrk 348	&	00h48m47.1s	&	+31d57m25s	&	$	1.328	\pm	0.179	$	&	EAP	&	AD	&	$	0.711	\pm	0.109	$	&	EAP	&	AD	&	$		<	0.319	$	&	EAP	&	UD	\\
MCG +05-03-013	&	00h51m35.0s	&	+29d24m05s	&	$	1.135	\pm	0.086	$	&	EAP	&	A	&	$	0.512	\pm	0.053	$	&	EAP	&	A	&	$	0.158	\pm	0.030	$	&	EAP	&	A	\\
Mrk 352	&	00h59m53.3s	&	+31d49m37s	&	$	0.127	\pm	0.025	$	&	EAP	&	A	&	$		<	0.078	$	&	EAP	&	U	&	$		<	0.049	$	&	PAP	&	U	\\
ESO 195-IG021 NED03	&	01h00m35.0s	&	-47d52m04s	&	$	0.415	\pm	0.024	$	&	TF	&	A	&	$	0.194	\pm	0.013	$	&	TF	&	A	&	$	0.072	\pm	0.011	$	&	TF	&	A	\\
MCG -07-03-007	&	01h05m26.8s	&	-42d12m58s	&	$	0.206	\pm	0.013	$	&	TF	&	A	&	$	0.094	\pm	0.009	$	&	TF	&	A	&	$		<	0.093	$	&	PAP	&	U	\\
2MASX J01064523+0638015	&	01h06m45.3s	&	+06d38m02s	&	$	0.036	\pm	0.006	$	&	TF	&	A	&	$		<	0.074	$	&	PAP	&	U	&	$		<	0.082	$	&	PAP	&	U	\\
\hline
\end{tabular}
\medskip
Note. -- \textit{Column 1:} Name of the source. \textit{Column 2:} Right ascension in J2000 coordinates. \textit{Column 3:} Declination in J2000 coordinates. \textit{Column 4:} 250 \um{} flux density and 1$\sigma$ uncertainty in units of Jy. 5$\sigma$ upper limits are given for undetected sources. \textit{Column 5:} Method used to extract the photometry. Either ``TF'' for timeline fitting, ``PAP'' for point sources aperture photometry, or ``EAP'' for extended source aperture photometry. See text for details on the differences. \textit{Column 6:} Flag for the photometry. ``A'' = source was detected at greater than 5$\sigma$. ``U'' = 5$\sigma$ upper limit. ``C'' = likely foreground cirrus contamination. ``d'' = nearby companion however unlikely to affect photometry. ``D'' = nearby companion that is likely strongly affecting photometry. \textit{Columns 7--9:} Same as Columns 4--6 except for the 350 \um{} waveband. \textit{Columns 10--12:} Same as Columns 4--6 except for the 500 \um{} waveband. Full version of the table available in online version.
\end{minipage}
\end{table}
\end{landscape}

\subsection{Comparison with \herschel{} Reference Survey}
The \herschel{} Reference Survey (HRS, \citet{Boselli:2010fj}) is a guaranteed time key project that surveyed 323 nearby ($15<D<25$ Mpc) galaxies using SPIRE to explore the dust content in early and late-type galaxies. Cross-correlating our sample with HRS, we found four sources (NGC 3227, NGC 4388, NGC 4941, and NGC 5273) that are a part of both. We compared the fluxes published in \citet{Ciesla:2012lq} to our own and find a mean ratio $F_{\rm{BAT}}/F_{\rm{HRS}}$ of 0.88, 0.92, and 0.87 for the 250, 350, and 500 \um{} wavebands respectively. 

However, there are several distinct differences between the HRS analysis and ours with the major difference being the beam area sizes. \citet{Ciesla:2012lq} used beam areas of 423, 751, and 1587 arcsec$^{2}$ compared with 469.7, 831.7, and 1793.5 arcsec$^{2}$ for our analysis. To correct for this, we multiplied the HRS fluxes for the galaxies by 423/469.7, 751/831.7, and 1587/1793.5 for the 250, 350, and 500 \um{} bands respectively.  After this correction the flux comparison ratios change to 0.97, 1.02, and 0.99. The remaining few percent differences we attribute to the differences in observing mode, map maker (Scanamorphos vs. naive map), data reduction, and photometry techniques. The ratios are also well within the uncertainties, therefore we conclude our photometry is consistent with the HRS.

\subsection{Comparison with \textit{Planck}}
We also compared our fluxes with those from the \textit{Planck} Catalog of Compact Sources \citep[PCCS;][]{Planck-Collaboration:2013rt}. The \textit{Planck} telescope performed an all-sky survey at nine submillimeter and radio wavebands to primarily measure the cosmic microwave background. The highest frequency band centered at 857 GHz matches the SPIRE 350 \um{} waveband and the 545 GHz (550 \um{}) overlaps the SPIRE 500 \um{} waveband allowing for independent measurements of the flux density of our sources. We searched the PCCS for our sources at each frequency using a 4' search radius and found 60 matches at 350 \um{} and 37 at 500 \um{}. To be consistent with our work we chose the aperture fluxes to compare with ours except for Centaurus A, NGC 1365, and M106 in which we chose the fluxes from fitting a Gaussian. These three sources are resolved even with \textit{Planck}'s poor spatial resolution, so the aperture fluxes will underestimate the true flux because the aperture sizes are equal to the resolution at each frequency. 

We applied color corrections to the \textit{Planck} fluxes to account for the differences in both central wavelength and spectral response. These were downloaded from the NHSC website\footnote{\url{https://nhscsci.ipac.caltech.edu/sc/index.php/Spire/PhotDataAnalysis}} and provides corrections for different temperature greybodies with an assumed emissivity of 1.8. Also provided are $F_{545}/F_{847}$ flux ratios that correspond to each temperature, which we compare with each observed flux ratio to find the right color correction for each source. Therefore we also restricted our comparison to only include sources that were detected in both the 545 and 857 GHz band giving a total of 27 sources. 

After correcting the \textit{Planck} fluxes, we compare them to our SPIRE fluxes and find a median SPIRE-to-\textit{Planck} ratio of 0.90 and 1.00 for the 350 and 500 \um{} band respectively. This shows a relatively good agreement between the SPIRE and \textit{Planck} instruments, especially in the 500 \um{} band and especially given all the assumptions and corrections that were made to compare the flux densities.

\section{FIR Properties of the \herschel-BAT Sample }
\subsection{Detection Rate and Luminosity Distributions}\label{sec:det_rate_lum_dist}
\citet{Melendez:2014yu} in analyzing the PACS photometry found 95\% and 83\% of the BAT sample had a 5$\sigma$ detection at 70 and 160 \um, indicating a largely complete survey of AGN for those wavelengths. Our SPIRE analysis finds a 5$\sigma$ completeness of 86\%, 72\%, and 46\% for 250, 350, and 500 \um{} respectively. The decreasing completeness reflects both the decreasing sensitivity of SPIRE with increasing wavelength as well as the rapid fall-off of the SED at longer wavelengths. Even with the relatively low detection rate at 500 \um, this still results in 143 AGN having complete FIR SEDs from 70--500 \um, representing a great step forward in advancing the study of the mid-far IR SED of AGN. 

\begin{figure}
\begin{center}
\includegraphics[width=\columnwidth]{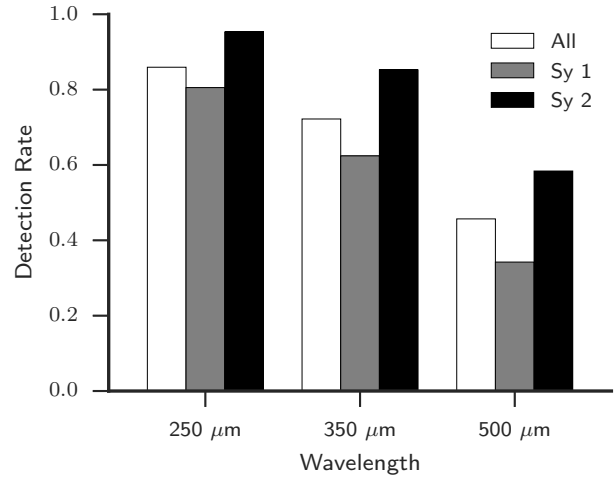}
\caption{\label{fig:det_frac} Detection fractions for the whole BAT sample (313 AGN), Sy 1's (139 AGN), and Sy 2's (169 AGN) in each SPIRE waveband. Sy 1's show significantly lower detection fractions than Sy 2's.}
\end{center}
\end{figure}

After splitting the sample into Sy 1's and Sy 2's, we find a distinct difference in the detection rate (Figure~\ref{fig:det_frac}). Sy 2's, for all 3 wavebands, are detected at a significantly higher rate than Sy 1's (95\% vs. 81\% for 250 \um, 85\% vs. 62\% for 350 \um, 58\% vs. 34\% for 500 \um).

\begin{figure}
\begin{center}
\includegraphics[width=\columnwidth]{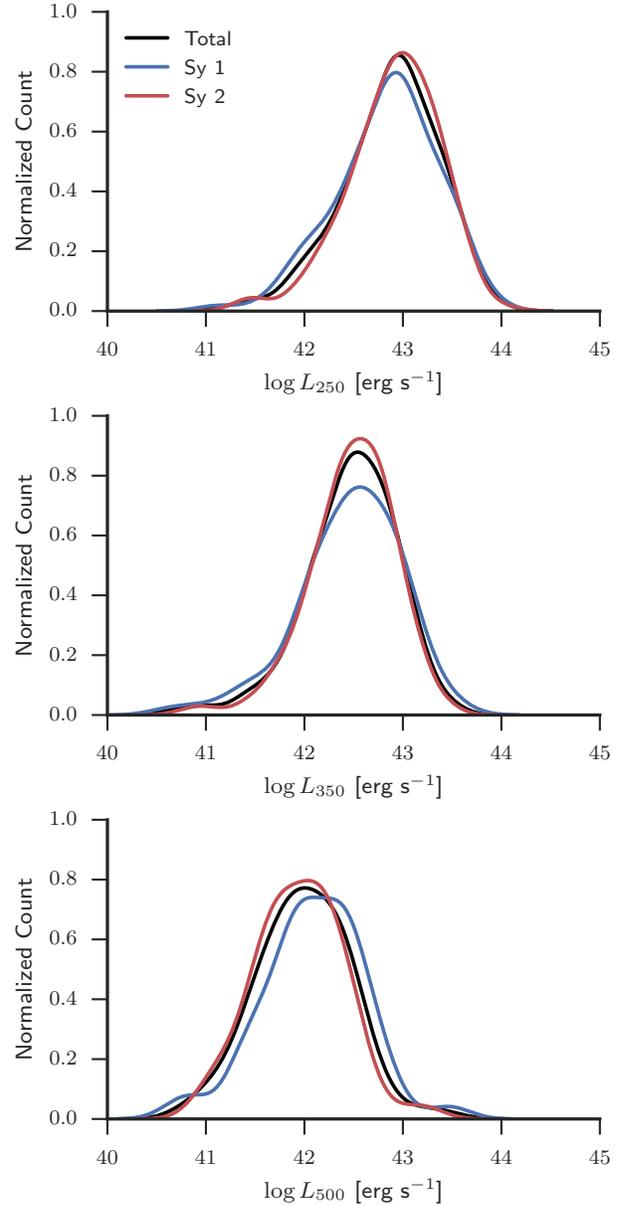}
\caption{\label{fig:lum_dist} Kernel density estimates (KDE, see text for explanation) of the luminosity distribution at each SPIRE wavelength for the total sample (black), Sy 1's (blue), and Sy 2's (red). At 250 \um{} Sy 1's and Sy 2's are nearly identical while at 350 and 500 \um{} Sy 2's seem to shift towards lower luminosities.}
\end{center}
\end{figure}

As we discuss in \citet{Melendez:2014yu} with the PACS photometry, this most likely is a result of the different redshift distribution between Sy 1's and Sy 2's. In the \textit{Swift}/BAT catalog, Sy 2's are found preferentially at lower redshifts than Sy 1's. Without taking into account the redshifts, the higher detection rate for Sy 2's would indicate Sy 2's favoring higher FIR luminosities. However, as Figure~\ref{fig:lum_dist} shows, this is not the case. Figure~\ref{fig:lum_dist} displays the kernel density estimates (KDE) of the SPIRE luminosity distributions for the total sample, Sy 1's, and Sy 2's. KDE's are a better way to represent distributions of data than histograms due to the histogram's dependence on both bin size and bin centers. A KDE represents each data point using a user-specified shape and sums all of them together. In this Paper, the shape we use is a Gaussian that has a width defined by ``Scott's Rule'' \citep{Scott:1992xy}.

Sy 1's and Sy 2's have identical 250 and 350 \um{} luminosity distributions. At 500 \um{} the luminosity distributions for Sy 2's actually peak at a \textit{lower} luminosity than Sy 1's. Again, this is due to Sy 2's occurring at lower redshifts where it is easier to detect the lower luminosities at longer wavelengths.

Figure~\ref{fig:lum_dist} however does not account for the undetected galaxies at each wavelength. For this, we use ``survival analysis'' to calculate the Kaplan--Meier product-limit estimator, a non-parametric representation of the survival function, $S(t) = P(T \geqq t)$. The Kaplan--Meier estimator then allows for an accurate estimate of mean, median, and standard deviation of a sample including censored values. 

To perform the survival analysis, we utilized the software package ASURV \citep{Feigelson:1985lr, Isobe:1990fk}. The only inputs necessary are the measured luminosities and upper limits for each wavelength. In Table~\ref{tab:mean_median_spire_lum} we outline the mean, 25th, 50th, and 75th percentiles of the luminosity distribution at each SPIRE wavelength. We give values for the entire sample as well as just the Sy 1's and Sy 2's.

\begin{table}
\begin{minipage}{6in}
\caption{SPIRE Luminosity Distributions}
\label{tab:mean_median_spire_lum}
\begin{tabular}{@{}c c c c c}
\hline
Sample & Mean & 25th \%tile & 50th \%tile & 75th \%tile\\
\hline
    \textbf{250 \um} &  &  &  &   \\ 
    Total & 42.8$\pm$0.03 & 42.5 & 42.9 & 43.2  \\ 
    Sy 1 & 42.6$\pm$0.07 & 42.2 & 42.8 & 43.1  \\ 
    Sy 2 & 42.8$\pm$0.05 & 42.6 & 42.9 & 43.2  \\ 
    \textbf{350 \um} &  &  &  &   \\ 
    Total & 42.3$\pm$0.04 & 42.0 & 42.4 & 42.7  \\ 
    Sy 1 & 41.9$\pm$0.1 & 40.9 & 42.2 & 42.6  \\ 
    Sy 2 & 42.3$\pm$0.05 & 42.0 & 42.4 & 42.7  \\ 
    \textbf{500 \um} &  &  &  &   \\ 
    Total & 41.6$\pm$0.06 & 41.3 & 41.7 & 42.0  \\ 
    Sy 1 & 41.2$\pm$0.1 & 26.7 & 41.1 & 41.9  \\ 
    Sy 2 & 41.4$\pm$0.09 & 40.6 & 41.6 & 42.0  \\ 
\hline
\end{tabular}
\end{minipage}
\end{table}
  
The mean and medians for Sy 1's and Sy 2's are similar, however Sy 2's do seem to have slightly higher luminosities at each wavelength. The mean luminosities deviate between 0.2--0.4 dex and the median deviates between 0.1--0.5 dex. We test for differences between the two samples using the Peto \& Prentice Generalized Wilcoxon test, which is similar to the standard Kolmogorov-Smirnov test but allows for censoring (i.e. upper limits). The test indicates that the probability that Sy 1's and Sy 2's are drawn from the same parent population is 5\%, 3\%, and 14\% for 250, 350, and 500 \um{} respectively. The usual cutoff for significant differences between two samples is 5\%, therefore we consider the luminosity distributions of Sy 1's and Sy 2's to be statistically the same at 500 \um{} and marginally different at 250 and 350 \um. This echoes the same small differences seen in \citet{Melendez:2014yu} for the 160 \um{} band where Sy 2's displayed slightly higher luminosities as well, suggesting that Sy 2's do indeed exhibit either larger dust masses or higher rates of star formation. 

\subsection{\herschel{} Undetected Sources}
With the photometry for all five wavebands of our \herschel{} study now measured, we can identify sources that are completely undetected at the 5$\sigma$ limit. In total, we find 11 sources that are completely undetected or about 4\%. These sources are 2MASS~J17485512-3254521, 2MASX~J08032736+0841523, 2MASX~J09360622-6548336, 2MASX~J12475784-5829599, 2MASX~J13512953-1813468, 2MASX~J14080674-3023537, 2MASX~J14530794+2554327, 2MASX J20183871+4041003, Arp 151, LEDA~138501, and PG~2304+042.

The images for three of these sources (2MASS J17485512-3254521, 2MASX~J12475784-5829599, and 2MASX~J20183871+4041003) are dominated by foreground cirrus emission so it is possible they would have been detected in the SPIRE wavebands without the cirrus contamination. The other eight \herschel-BAT AGN, however, are clear non-detections at all five wavebands. 

\begin{figure}
\includegraphics[width=\columnwidth]{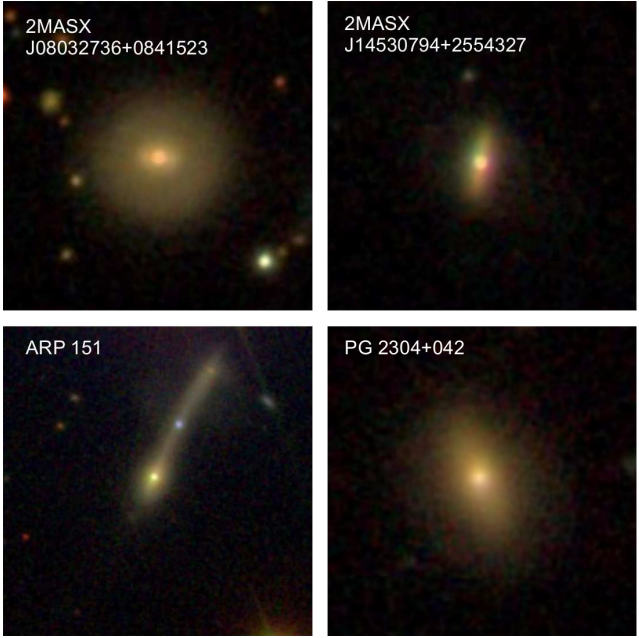}
\caption{\label{fig:undetected} SDSS \emph{gri} images of 4/8 the \herschel-BAT AGN that are undetected at all five \herschel\ wavebands.}
\end{figure}

We examined SDSS optical images (Figure~\ref{fig:undetected}) for four of the eight sources that are not detected in any \herschel{} waveband. 3/4 of the host galaxies are appear to be quite red in color, indicating a lack of young stars and older stellar population. Further their morphologies are either elliptical or cigar shaped, indicative of early-type galaxies which are known to be faint in the FIR and contain little dust \citep[e.g.][]{2013A&A...552A...8D}. The only exception is ARP~151, which seems to be involved in a merger, displaying a long and narrow tidal tail. It is possible the merger process has removed large amounts of gas and dust from the galaxy causing it to be faint in the FIR. Given 75\% of the undetected sources with optical images are early-type galaxies, it is likely the remaining four AGN host galaxies are also early-type galaxies with relatively little cold dust and star formation occurring. 

\subsection{Wavelength--Wavelength Luminosity Correlations}
Thermally heated dust is thought to produce the mid-far IR SED \citep[e.g.][]{Draine:2003gd}. Shorter wavelength emission corresponds to both hotter and smaller dust grains and vice versa for longer wavelengths. Both the amount of dust (i.e. dust mass) heated to a specific temperature as well as the relative intensity of the heating process determine the strength of the emission at a particular wavelength. If the same process (e.g. young star formation) is heating all of the dust and producing the entire FIR SED, we would expect strong correlations between each wavelength, however if two or more disjointed processes contribute to the SED (e.g. star formation and AGN emission), correlations will become weaker.

Three processes could contribute to the heating of dust in the BAT AGN. Recent star formation in the galaxy will produce OB stars with a high intensity of UV light that can heat nearby dust to large temperatures. UV light can also escape the star-forming regions and heat dust further away to colder temperatures. Older stellar populations, however, also produce an interstellar radiation field that can heat diffuse dust to temperatures around 15 K which would contribute most heavily at the longest wavelengths \citep[e.g.][]{Bendo:2010kq, Bendo:2012lr, Boselli:2012qv, Bendo:2015lr}. Finally, the UV light from the AGN itself can heat dust in the torus.

We ran a correlation analysis between each \textit{Herschel} waveband. Two effects must be taken into account to establish reliable correlation coefficients: censoring and confounding variables. The confounding variable in this case is distance. Since our sample is flux-limited, higher luminosity objects are more likely to be found at larger distances. Therefore it can produce the effect of an intrinsic correlation when comparing two luminosities. To mitigate the effects of censoring and the luminosity-distance relationship, we calculated the partial Kendall-$\tau$ correlation coefficient as presented in \citet{Akritas_1996}. Table~\ref{tab:wave_corrs} displays all of the correlation coefficients ($\rho_{\tau}$) as well as the probability of zero correlation ($P_{\tau}$).

\begin{table*}
\begin{minipage}{4.25in}
\caption{Luminosity Partial Correlation Coefficients}
\label{tab:wave_corrs}
\begin{tabular}{@{}c c c c}
\hline
 & 250 \um & 350 \um & 500 \um\\
 \hline
    \textbf{Total} & & & \\ 
    70 \um & 0.54$\pm$0.03 ($\ll$ 0.01) & 0.45$\pm$0.03 ($\ll$ 0.01) & 0.27$\pm$0.03 ($\ll$ 0.01) \\ 
    160 \um & 0.74$\pm$0.02 ($\ll$ 0.01) & 0.62$\pm$0.02 ($\ll$ 0.01) &  0.34$\pm$0.03 ($\ll$ 0.01)\\ 
    250 \um & ... & 0.75$\pm$0.02 ($\ll$ 0.01) &  0.41$\pm$0.03 ($\ll$ 0.01)\\ 
    350 \um & ... & ... & 0.42$\pm$0.03 ($\ll$ 0.01) \\
    14--195 keV & 0.06$\pm$0.03 (0.05)  & 0.06$\pm$0.03 (0.06) & 0.02$\pm$0.02 (0.52) \\
    &&& \\
    \textbf{Sy 1} &&&\\
    70 \um & 0.55$\pm$0.04 ($\ll$ 0.01) & 0.43$\pm$0.04 ($\ll$ 0.01) & 0.23$\pm$0.04 ($\ll$ 0.01) \\
    160 \um & 0.71$\pm$0.03 ($\ll$ 0.01) & 0.57$\pm$0.04 ($\ll$ 0.01) & 0.29$\pm$0.04 ($\ll$ 0.01) \\
    250 \um & ... & 0.66$\pm$0.04 ($\ll$ 0.01) &  0.32$\pm$0.04 ($\ll$ 0.01)\\
    350 \um & ... & ... & 0.32$\pm$0.04 ($\ll$ 0.01) \\
    14--195 keV & 0.13$\pm$0.05 (0.003)  & 0.10$\pm$0.04 (0.02) & 0.04$\pm$0.04 (0.23) \\
    &&& \\
    \textbf{Sy 2} &&&\\
    70 \um & 0.51$\pm$0.04 ($\ll$ 0.01) & 0.44$\pm$0.04 ($\ll$ 0.01) & 0.30$\pm$0.04 ($\ll$ 0.01) \\
    160 \um & 0.74$\pm$0.03 ($\ll$ 0.01) & 0.64$\pm$0.03 ($\ll$ 0.01) & 0.40$\pm$0.04 ($\ll$ 0.01) \\
    250 \um & ... & 0.81$\pm$0.03 ($\ll$ 0.01) &  0.48$\pm$0.04 ($\ll$ 0.01)\\
    350 \um & ... & ... & 0.50$\pm$0.04 ($\ll$ 0.01) \\
    14--195 keV & 0.02$\pm$0.04 (0.69)  & 0.03$\pm$0.05 (0.54) & -0.004$\pm$0.04 (0.91) \\
\hline
\end{tabular}
\end{minipage}
\end{table*}

While all the relationships show some amount of correlation with very low ($\ll$ 1\%) probabilities of occurring by chance, the strongest ones occur between wavelengths that are nearest each other. The 160 vs. 250 \um{} and 250 vs. 350 \um{} correlations have a correlation coefficient $>0.7$. This makes sense within the context of multiple temperature components. Photometry from nearby wavelengths should be produced from closely related temperature components.

The weak correlation between 70 and 500 \um{} indicates the emission in these wavebands does not originate from closely related processes. 70 \um{} emission comes from much hotter and smaller dust grains than 500 \um{} and several processes could provide an explanation. Since this is an AGN sample, there could be a strong contribution from AGN heated dust at 70 \um{}, whereas at 500 \um{}, AGN related emission would likely be negligible. This is supported by our findings in \citet{Melendez:2014yu} where we showed that the 70 \um{} luminosity is weakly correlated with AGN luminosity. Further, in \citet{Mushotzky:2014ad} we found that the BAT AGN morphologies at 70 \um{} were concentrated in the nucleus potentially indicating an AGN contribution.

The weak correlation, however, can also be explained if non-star-forming processes also contribute to the 500 \um{} emission. While in non-AGN galaxies, the majority of 70 \um{} emission is most likely due to small, stochastically heated dust grains around HII regions, $>250$ \um{} emission is likely produced by the heating of larger dust grains in the diffuse ISM by older stars \citep[e.g.][]{Bendo:2015lr}. Therefore, the disconnect between the stellar populations would produce significant scatter in the correlation between 70 and 500 \um. 

A third possibility is that synchrotron radiation produced by radio jets associated with AGN can contribute to the FIR, especially the longest wavelengths as seen in some radio-loud galaxies \citep{Baes:2010ek,Boselli:2010fr}. This non-thermal emission would be completely unrelated to the thermal emission at 70 \um, thereby producing a weaker correlation between the luminosities at those wavebands. In a later section we will show there are indeed some radio-loud sources in our sample where synchrotron emission dominates the SPIRE emission, although the fraction of sources is quite low. 

When we break the sample down into Sy 1's and 2's we do not find much difference between the correlation coefficients. This shows that Sy 1's and 2's are not different in terms of their overall FIR emission and the same processes are likely producing the FIR emission. Sy 1's do show a slightly weaker correlation between the \textit{Herschel} luminosities especially the ones involving 500 \um. This is likely due to the fact that most radio-loud AGN are classified as Sy 1's so synchrotron emission is contributing strongest at 500 \um{} compared to the other wavebands.

\begin{figure}
\begin{center}
\includegraphics[width=\columnwidth]{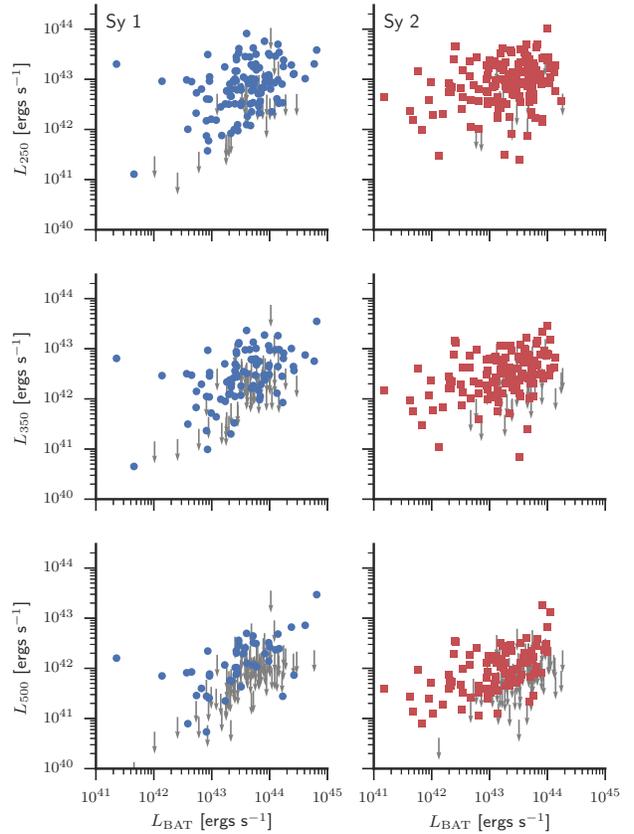}
\caption{\label{fig:lum_spire_BAT} Correlations between each SPIRE waveband luminosity and the BAT 14--195 keV luminosity. Blue circles in the left column represent Sy 1's. Red squares in the right column are Sy 2's. Sources with gray arrows indicate $5\sigma$ upper limits.}
\end{center}
\end{figure}

\subsection{Correlation With Ultra-Hard X-ray Luminosity}
Ultra-hard X-ray luminosity directly probes the current strength of the AGN because it likely originates very close to the SMBH. The 14--195 keV luminosity then provides an unambiguous measure of the AGN power especially for Compton-thin sources. If we want to determine whether the AGN contributes in any way to the FIR luminosity, the first check would be to correlate the 14--195 keV luminosity with each waveband's luminosity. \citet{Melendez:2014yu} ran correlation tests for the PACS wavebands finding a weak, but statistically significant correlation between the 70 and 160 \um{} luminosity and the 14--195 keV luminosity for Sy 1's but not for Sy 2's.

Using the same methods as we did to measure strengths of the correlations between each \textit{Herschel} luminosity, we measured the correlation between each SPIRE and 14--195 keV luminosity. The last lines of each section of Table~\ref{tab:wave_corrs} lists the results of the correlation tests and Figure~\ref{fig:lum_spire_BAT} plots the correlations with gray arrows indicating upper limits.

For the AGN sample as a whole, no significant correlation exists between the SPIRE and 14--195 keV luminosity. All of the $\rho_{\tau}$, after accounting for the partial correlation with distance, are below 0.1 with $P_{\tau}$ either at or above 5\%. However, when we break the sample up into Sy 1's and Sy 2's and redo the correlation tests, we find a very weak correlation between the 250 and 350 \um{} luminosity and ultra-hard X-ray luminosity for Sy 1's only ($\rho_{\tau}=0.13$ and 0.10). Sy 2's $\rho_{\tau}$ are consistent with no correlation with $P_{\tau}>54\%$ for all three wavebands. This continues the trend with what was found in \citet{Melendez:2014yu} where only Sy 1's were found to have a weak correlation between the BAT luminosity and the PACS waveband luminosities. The partial correlation coefficients were $0.20\pm0.04$ and $0.12\pm0.04$ for Sy 1s and $0.08\pm0.04$ and $-0.005\pm0.04$ for Sy 2's at 70 and 160 \um{} respectively \citep[see Table 3 of][]{Melendez:2014yu}\textbf{}. 

We note however that except for the 70 \um{} waveband, none of the correlation coefficients are $>3\sigma$ away from a null correlation coefficient. So even though $P_{\tau}< 5\%$, these are all quite weak correlations between the \herschel{} wavebands and BAT luminosity for Sy 1's. At 500 \um{}, the correlation completely disappears. 

As we discuss in \citet{Melendez:2014yu}, this extends the trend observed in the MIR where strong correlations have been measured between the 9, 12, and 18 \um{} luminosities and the BAT luminosity \citep{Gandhi:2009kx, Matsuta:2012gf, Ichikawa:2012ul} but moving towards longer wavelengths the correlation degrades rapidly as shown in \citet{Ichikawa:2012ul} for 90 \um{} emission. 

Clearly then, at long wavelengths ($\lambda > 40$ \um), emission from dust unrelated to the AGN dominates most galaxies. However, we must still explain why Sy 1's retain a weak correlation while Sy 2's do not. \citet{Melendez:2014yu} discussed in detail several theories for why Sy 1's would show a different correlation between the \herschel{} luminosities and BAT luminosity. These included an intrinsically different BAT luminosity distribution for Sy 1's and Sy 2's and the addition of Compton-thick (CT) AGN in the Sy 2 sample.

Several authors have found that the Sy 2 luminosity function breaks at a significantly lower luminosity than for Sy 1s \citep[e.g.][]{Cowie:2003fk, Hasinger:2005qy, Burlon:2011pi}. At low BAT luminosity, then, there are more Sy 2's than Sy 1's as is evident in Figure~\ref{fig:lum_spire_BAT}. \citet{Rosario:2012fr} showed that at low AGN luminosity the correlation between SFR and AGN luminosity flattens. This can be explained one of two ways: 1.) Only at high AGN luminosity is there a direct connection between star formation and AGN activity. 2.) At high AGN luminosity, the IR-related AGN emission overwhelms any star-forming related IR emission even at long wavelengths. Regardless of the physical reason, the flattening of the SF-AGN relationship at low luminosity could explain the correlation differences seen between Sy 1s and Sy 2s since Sy 2s are preferentially found at lower luminosity than Sy 1s. 

\citet{Melendez:2014yu} tested this for the PACS wavebands and found that only using high luminosity objects did not improve the X-ray-to-IR correlation for Sy 2's. We repeated this test with the SPIRE luminosities and limited the samples to only AGN with BAT luminosity greater than $10^{43.5}$ ergs s$^{-1}$. For both Sy 1's and Sy 2's the correlations become insignificant, likely because of the reduction in number of sources used in the analysis. It is then inconclusive whether or not a difference in intrinsic AGN luminosity is the cause of the differences in correlations between X-ray and IR luminosity for Sy 1's and Sy 2's.


The other possibility is that CT sources are contaminating the Sy 2 sample. This would have an effect if the high column density ($N_{\rm H} > 10^{24}$ cm$^{-2}$) material obscuring the AGN scatters 14-195 keV photons out of our line sight causing a lower measured BAT luminosity. \citet{Melendez:2014yu} identified 44 either confirmed CT AGN or likely CT AGN based on X-ray hardness ratios in our sample. We removed these likely CT sources and redid the correlation tests, finding no difference from before just as \citet{Melendez:2014yu} found. Therefore, it is unlikely that CT sources are the cause of the difference between the Sy 1 and Sy 2 correlations.

Given the inconclusiveness of the first test limiting the sample to high luminosity objects, we can only speculate about the reason for the difference in correlations. However, \citet{Melendez:2014yu} did find that restricting the sample to high luminosity objects increased the strength of the correlation for Sy 1's but not Sy 2's in the PACS wavebands. It is possible then that either a direct physical link between the SFR and AGN luminosity that is only evident in high luminosity AGN or increased contamination of the AGN to the IR SED is causing the relatively stronger correlation in Sy 1s but not Sy 2s.


What is conclusive is that the SPIRE emission from the AGN host galaxies on average is not strongly contaminated by AGN-related emission given the small values for the correlation coefficients even for Sy 1's. 

\subsection{SPIRE Colors}
While in the previous sections, we examined the absolute luminosities of each SPIRE waveband and the correlations between each other and other wavebands (PACS and BAT), in this section we examine the SPIRE colors (i.e. flux ratios). Colors in general provide measures of the shape of the SED. Different objects and mechanisms produce significantly different SED shapes across the same wavelength regime, therefore colors can be used to separate distinct populations from each other especially when groups display the same absolute brightnesses. We investigate two colors, $F_{250}/F_{350}$ and $F_{350}/F_{500}$, that probe the Rayleigh Jeans tail of a modified blackbody if the dominant process producing the emission is cold dust. 

\subsubsection{BAT AGN SPIRE colors are similar to high-mass non-AGN galaxies}
\begin{figure*}
\begin{center}
\includegraphics[width=0.8\textwidth]{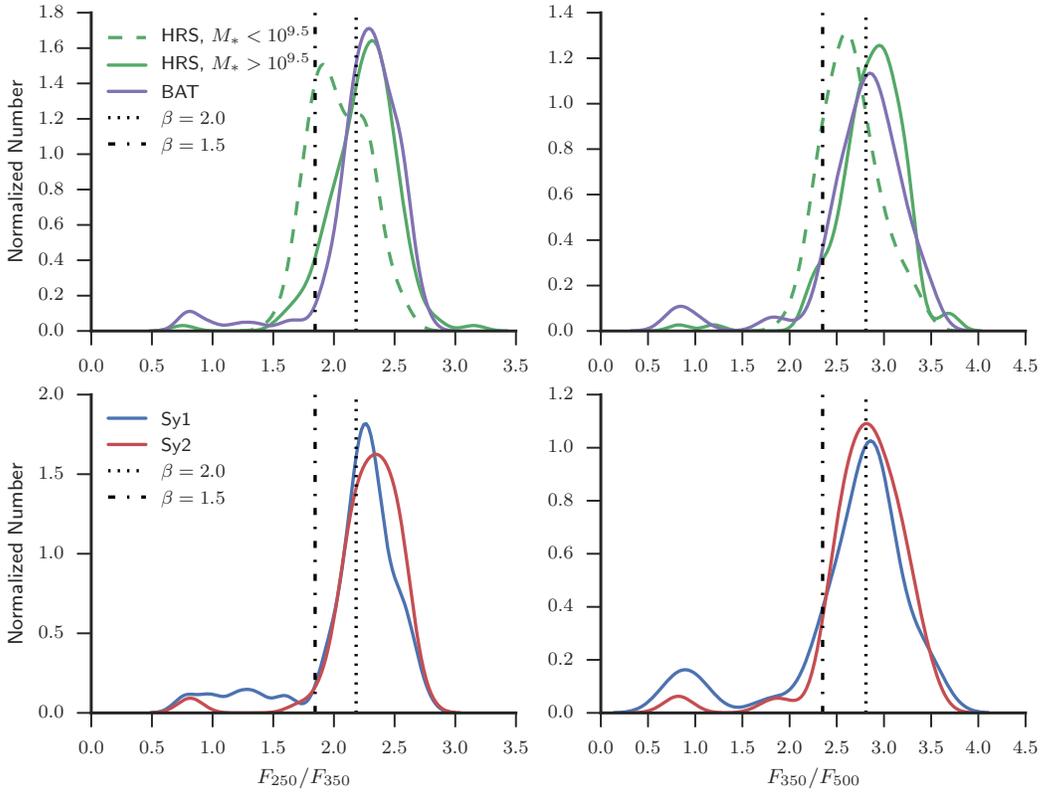}
\caption{\label{fig:hist_colors} \textit{top row:} KDE of the  $F_{250}/F_{350}$ (left) and  $F_{350}/F_{500}$ (right) colors for the BAT AGN (purple) and HRS (green). For the HRS galaxies, we split the sample into high ($M_{*} > 10^{9.5}$ M$_{\odot}$; solid line) and low ($M_{*} < 10^{9.5}$ M$_{\odot}$, dashed linje) stellar mass groups. A K-S test indicates the SPIRE color distributions for the BAT AGN and HRS high-mass group are statistically the same with $P_{K-S} = 49\%$ for $F_{250}/F_{350}$ and $P_{K-S} = 22\%$ for $F_{350}/F_{500}$. \textit{bottom row:} KDEs for the colors of the BAT AGN separated into Sy 1s (blue) and Sy 2s (red). K-S tests indicate the two Seyfert types are drawn from the same parent population with $P_{K-S} = 17\%$ and $37\%$ for $F_{250}/F_{350}$ and $F_{350}/F_{500}$ respectively. In both rows we also plot the expected color for a modified blackbody with a dust temperature of 20 K and an emissivity of 2.0. (dotted line) and 1.5 (dot-dash line)}
\end{center}
\end{figure*}

Figure~\ref{fig:hist_colors} plots the KDE of the two colors. The top row compares the distribution of the colors ($F_{250}/F_{350}$ on the left and $F_{350}/F_{500}$ on the right) from the BAT AGN and HRS samples. While the HRS galaxies are local like the BAT AGN, one major difference is the stellar mass distribution. The HRS sample contains more low stellar mass galaxies while the BAT AGN are strictly found in galaxies with stellar mass ($M_{*}$) values above $10^{9.5}$ M$_{\odot}$ \citep{Koss:2011vn}. 

As \citet{Boselli:2012qv} show, FIR colors can be affected by the physical properties of the galaxy, especially the colors probing the cold dust such as the ones we are investigating here. Therefore, we broke the HRS sample into two groups, a high mass group ($M_{*} >  10^{9.5}$ M$_{\odot}$) and low mass one ($M_{*} >  10^{9.5}$ M$_{\odot}$) indicated in Figure~\ref{fig:hist_colors} by the solid and dashed green lines. Stellar masses for the HRS were obtained from \citet{Cortese:2012fj}.

We also plot the theoretical color of the modified blackbody with a dust temperature of 20 K and emissivity ($\beta$) of 2.0 and 1.5, values typical of normal, star-forming galaxies \citep[e.g.][]{Calzetti:2000fk, Smith:2012fj, Galametz:2012uq, Dale:2012dq, Cortese:2014qq}. The HRS high mass group and BAT AGN display nearly identical color distributions for both colors whereas the HRS low mass group is skewed toward lower colors. Results of a K-S test show that the HRS high mass group and BAT AGN colors are drawn from the same parent population with a $P_{K-S} = 49\%$ and 22\% for $F_{250}/F_{350}$ and $F_{350}/F_{500}$ respectively. On the other hand the HRS low mass group colors are significantly different from the BAT AGN with $P_{K-S}$  values much less than 1\%. 

This is consistent with what was found in \citet{Boselli:2012qv}, who showed that the SPIRE colors for the HRS sample were affected by the metallicity of the galaxy with metal rich galaxies displaying larger flux ratios and a higher $\beta$ than metal poor ones. Given the strong, positive relationship between metallicity and $M_{*}$ \citep[e.g.][]{Tremonti:2004fq}, this is exactly in line with what is seen in Figure~\ref{fig:hist_colors}. The HRS high mass group and BAT AGN display colors closer to the ones expected for a modified blackbody with $\beta\sim2.0$ while the low mass HRS group are closer to $\beta\sim1.5$. 

\begin{figure*}
\begin{center}
\includegraphics[width=0.8\textwidth]{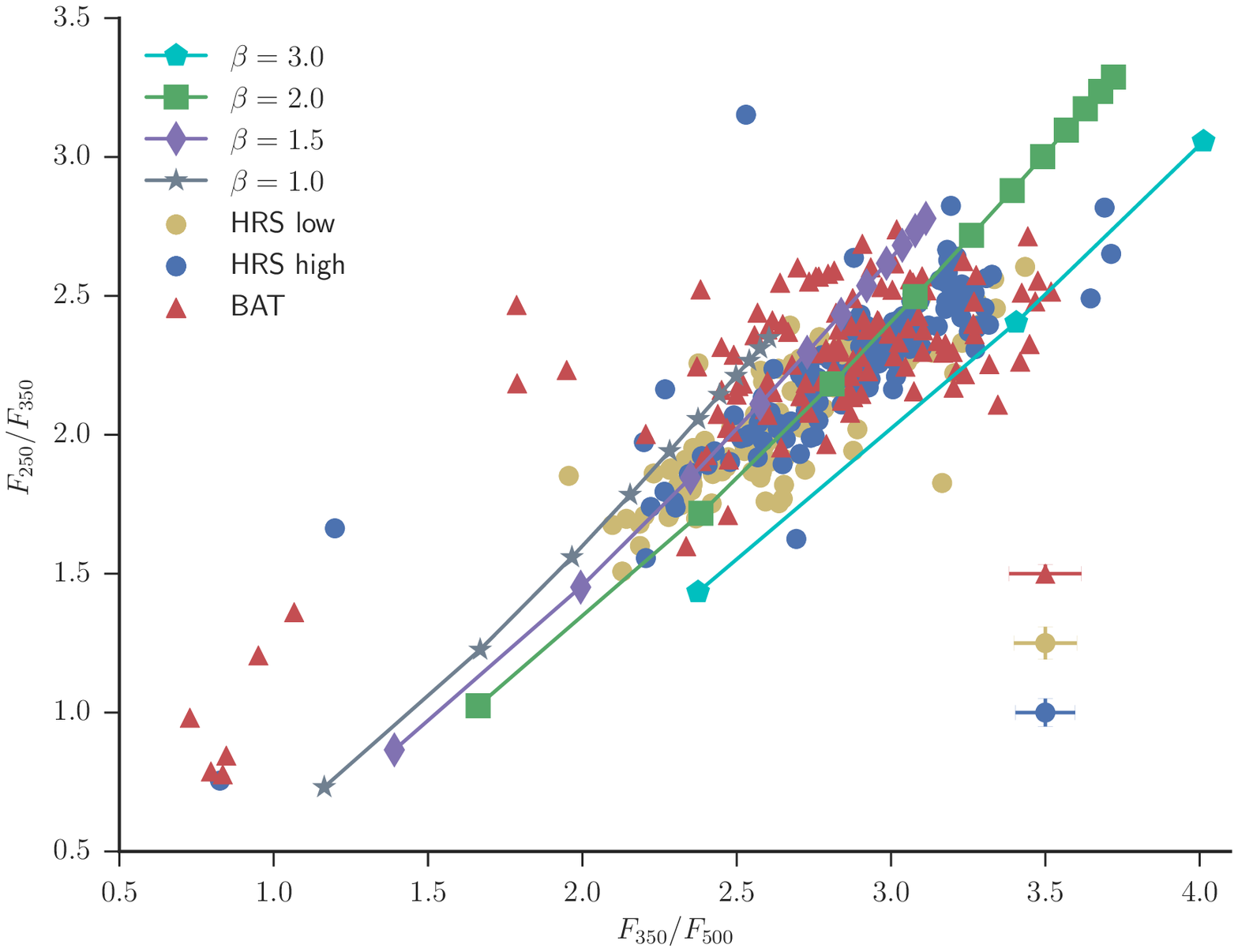}
\caption{\label{fig:color-color} SPIRE color-color plot of the BAT AGN (red triangles) and HRS high and low $M_{*}$ group (blue and yellow circles). The lines with markers represent theoretical colors assuming a modified blackbody ($F_{\nu} \propto \nu^{\beta}B(\nu, T)$) with emissivity, $\beta=3.0, 2.0, 1.5, \mathrm{and}\, 1.0$. Each marker represents a different temperature from 10 K (lower left) to 60 K (upper right) in increments of 5 K, except for $\beta=3.0$ for which we only show markers for 10, 15, and 20 K.  The main locus for both the BAT AGN and HRS is concentrated around the theoretical colors between 15--30 K and $\beta=1.5-2.0$. Representative error bars are shown in the lower right corner.}
\end{center}
\end{figure*}

In Figure~\ref{fig:color-color} we plot both colors together for the HRS and BAT AGN. Nearly all of the HRS galaxies are concentrated along a main locus as well as many of the BAT AGN. We also plot the expected colors for a modified blackbody with varying temperature between 10 and 60 K and an emissivity of either 2.0 (green line and squares) or 1.5 (purple line and diamonds). Each square or diamond represents an increase of 5 K starting at 10 K in the lower left. The main locus for both samples is clearly aligned with a modified blackbody with temperatures between 15--30 K. \cite{Cortese:2014qq} fit the FIR SED of the HRS sample using a single temperature modified blackbody finding exactly this range of temperatures and an average emissivity of 1.8. Further these values are consistent with dust in the Milky Way, Andromeda, and other nearby galaxies \citep{Galametz:2012uq, Boselli:2012qv, Smith:2012fj}.

\subsubsection{Sy 1s and Sy 2s show the same SPIRE colors}
In the bottom rows, we compare Sy 1's and Sy 2's. Based on the results of our analysis in Section~\ref{sec:det_rate_lum_dist}, we would expect Sy 1's and Sy 2's to show the same distribution of colors. Indeed this is the case as both distributions in both colors peak at nearly the same values and have nearly the same spread. K-S tests reveal the colors for the two Seyfert types are drawn from the same parent population with $P_{K-S} = 0.2$ for $F_{250}/F_{350}$ and $P_{K-S} = 0.3$ for $F_{350}/F_{500}$.

\subsubsection{Outliers in SPIRE color-color space: Radio-loud AGN and excess 500 \um{} emission}
While the bulk of the SPIRE colors are very similar between the HRS and BAT, and the two Seyfert types, one noticeable difference is a distinct bump in the color distribution around 0.75. This bump is absent in the HRS sample and mainly is made up of Sy 1s. With both flux ratios less than one, this indicates a monotonically rising SED that is in stark contrast with the rapidly declining SED characteristic of a modified blackbody. The equation for a modified blackbody is
\begin{equation}\label{eq:mod_blackbody}
F_{\nu} \propto \nu^{\beta}B(\nu, T)
\end{equation}
\noindent where $B(\nu, T)$ is the standard Planck blackbody function with a temperature of $T$.  

The bump seen in Figure~\ref{fig:hist_colors} is very evident in Figure~\ref{fig:color-color} as a separate population in the lower left-hand corner. Specifically 6 BAT AGN and one HRS galaxy occupy the region of color-color space where $F_{250}/F_{350} < 1.5$ and $F_{350}/F_{500} < 1.5$. Based on the theoretical curves, these exceptional colors cannot be explained as either a different temperature or emissivity. Rather an entirely different process is producing the FIR emission in these galaxies and since the colors indicate essentially a rising SED, we suspected synchrotron radiation as the likely emission mechanism with its characteristic increasing power law shape with wavelength.

Further there seems to be a horizontal spread in the distribution of the BAT AGN in Figure~\ref{fig:color-color} that is clearly not evident in the HRS. Also this effect is not seen Figure~\ref{fig:hist_colors} and the KDEs because it only becomes evident when analyzing the two colors together. Both samples span the same range of colors, however their distribution in color-color space is different. This is characterized by a large group of BAT AGN above and to the left of the main locus and $\beta=1.5$ line (purple) as well as a smaller group of AGN below and to the right of the main locus and $\beta=2.0$ line (green). The latter group can be explained simply from a decrease in temperature and increase in emissivity up to a beta value of 3.0 (cyan line in Figure~\ref{fig:hist_colors}), indicating the prevalence of large amounts of cold dust. The former group could be explained by a decrease in the emissivity closer to around values of 1.0 (gray line), however this would require the dust temperature to increase to values above 60 K, not typical of regular star-forming galaxies.

Rather these high temperatures (70--100 K) are near the expected temperatures for dust heated by the AGN, which show characteristic peaks in their SED between 20--40 \um{} \citep{Richards:2006fj,Netzer:2007ve,Mullaney:2011yq}. If the AGN is affecting the colors of these sources more than the ones on the main locus then there should be some correlation between the offset from the main locus and an indicator of AGN strength such as X-ray luminosity. 

To quantify the offset from the main locus, we fit the SED of all of the sources in Figure~\ref{fig:color-color} using a modified blackbody (Eq.~\ref{eq:mod_blackbody}) with a fixed emissivity of 2.0 to measure the excess or deficiency of observed 500 \um{} emission compared to the model. 

With the emissivity fixed at 2.0, there are only two free parameters, the dust temperature and normalization. We fit the sources within a Bayesian framework using uniform priors for the logarithm of the normalization and dust temperature and a standard Gaussian likelihood function. To sample the posterior probability density function, we use the \texttt{emcee}\footnote{Available at \url{http://dan.iel.fm/emcee/current/}} package \citep{Foreman_Mackey_2013} that implements the affine-invariant ensemble sampler for Markov chain Monte Carlo (MCMC) originally proposed by \citet{Goodman_2010}\footnote{The MCMC ensemble sampler is essentially multiple MCMC chains running in parallel and each chain is called a ``walker''. We chose to use 50 walkers that run for 1000 steps each. The first 200 steps of each walker are discarded as a ``burn-in'' period that allows each walker time to move away from the initial guesses for the parameters and begin exploring the full posterior probability distribution.}.

For the model fitting, we only use 160, 250, and 350 \um{} flux densities. We exclude the 500 \um{} data point because our aim is to compare the expected 500 \um{} emission from the model with the observed one and do not want the fitting influenced by the observed emission. We also exclude the 70 \um{} flux density because it can be dominated by emission from hotter dust heated by young stars in dense star-forming regions or the AGN itself \citep{Calzetti:2000fk,Bendo:2010kq,Boquien:2011qf,Smith:2012fj,Melendez:2014yu}.

\begin{figure*}
\begin{center}
\includegraphics[width=0.8\textwidth]{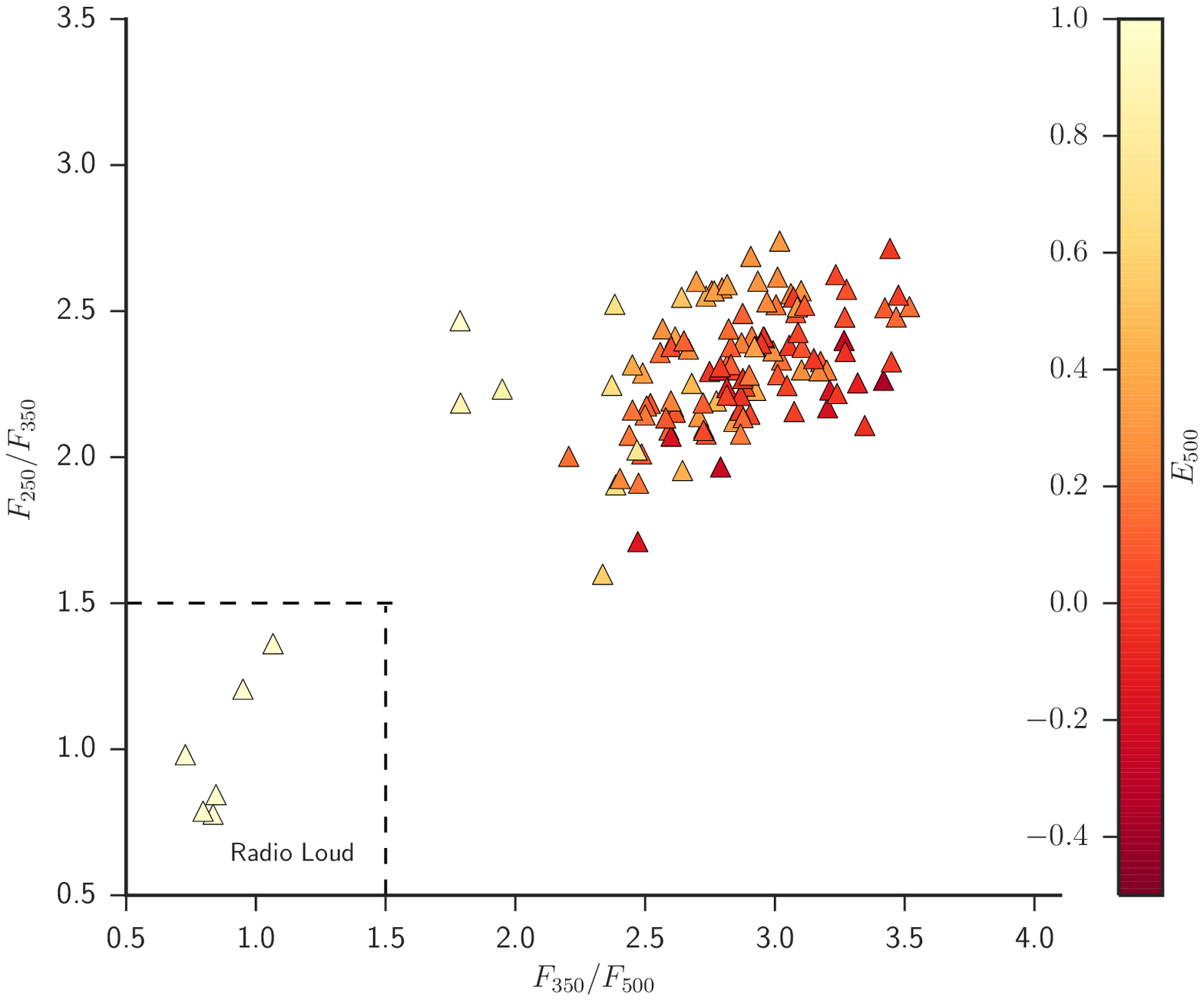}
\caption{\label{fig:color-color_excess} Same as Figure~\ref{fig:color-color} but only showing the BAT AGN. The points are colored by $F_{excess}$, the measured 500 \um{} excess emission compared to a modified blackbody. The two dashed lines delineate the color cutoff for radio loud AGN, $F_{250}/F_{350} < 1.5$ and $F_{350}/F_{500} < 1.5$.}
\end{center}
\end{figure*}

Each sample from the MCMC chain contains values for the parameters of the modified blackbody that are likely given the posterior distribution. From all of these parameters, we calculated 40000 modeled 500 \um{} emission and ``excess'' using the following equation:
\begin{equation}\label{eq:500um_excess}
E_{500} = \frac{F_{obs} - F_{model}}{F_{model}}
\end{equation}
\noindent $E_{500}$ then represents a fractional excess (or deficiency) as compared to the model emission. A deficiency would be indicated by a negative value for $E_{500}$. The final excess value associated with the source is then determined as the median of all of the excess values. In Figure~\ref{fig:color-color_excess} we plot the same color-color diagram as in Figure~\ref{fig:color-color} with each point colored by its measured $E_{500}$.

In general, points with low values of the $F_{350}/F_{500}$ color show high values of $E_{500}$ and vice versa for high values of the $F_{350}/F_{500}$ color. Points along the main locus are scattered around $E_{500} = 0$. Thus, $E_{500}$ can quantify a source's distance from the main locus and allows us to study possible causes for this excess emission at 500 \um.

We first measure the correlation between $E_{500}$ and radio loudness. AGN historically have been classified into two groups based on how bright their radio emission is compared to another waveband, usually optical. These groups are ``radio-loud'' and ``radio-quiet'' AGN with the former group showing bright radio emission and the latter faint radio emission relative to the optical or X-ray emission \citep{Kellermann:1989sf,Xu:1999ty}. While originally radio-loud and radio-quiet AGN seemed to form a dichotomy, the consensus now seems to be that there is a broad distribution of radio-loudness rather than a bimodality \citep{White:2000rz,Cirasuolo:2003rm,Cirasuolo:2003zl,Laor:2003yg}. Further, the original radio loudness parameter, $R = L_{\rm radio}/L_{\rm opt}$ which measured the ratio of the radio to optical luminosity, was shown to underestimate the radio loudness especially for low-luminosity Seyfert galaxies \citep{Terashima:2003fv}. Rather $R_{\rm X} = L_{\rm radio}/L_{\rm X}$ which measures the nuclear radio to X-ray luminosity ratio was confirmed to be a better radio-loudness indicator given X-rays are less affected by obscuration and contamination from the host galaxy. Therefore, for the BAT AGN, we use $R_{\rm X}$ to measure the radio-loudness with $L_{\rm radio} = L_{\rm 1.4\,GHz}$ and $L_{\rm X} = L_{\rm 14-195\,keV}$. 

For $L_{\rm 1.4\,GHz}$ we first cross-correlated the BAT AGN with the FIRST and NVSS databases which provide 1.4 GHz flux densities over all of the northern sky. FIRST flux densities were preferred over NVSS due to the much better angular resolution (5" vs. 45"). Since \swift/BAT was an all-sky survey, nearly half of the BAT AGN were not included in either FIRST or NVSS. For these southern sources we turned to the Sydney University Molonglo Sky Survey \citep[SUMSS;][]{Bock:1999fp} which surveyed the southern sky at 843 MHz. Finally, for the remaining sources missing radio data, we performed a literature search and found 5 GHz fluxes from various other studies \citep{Becker:1991qd,Griffith:1993qr,Rush:1996db,Ho:2001hl,Shi:2005rc}. To convert all flux densities to 1.4 GHz, we assumed a power-law spectrum, $F_{\nu} \propto \nu^{-0.7}$, that is typical for synchrotron emission. 

\begin{figure*}
\begin{center}
\includegraphics[width=\textwidth]{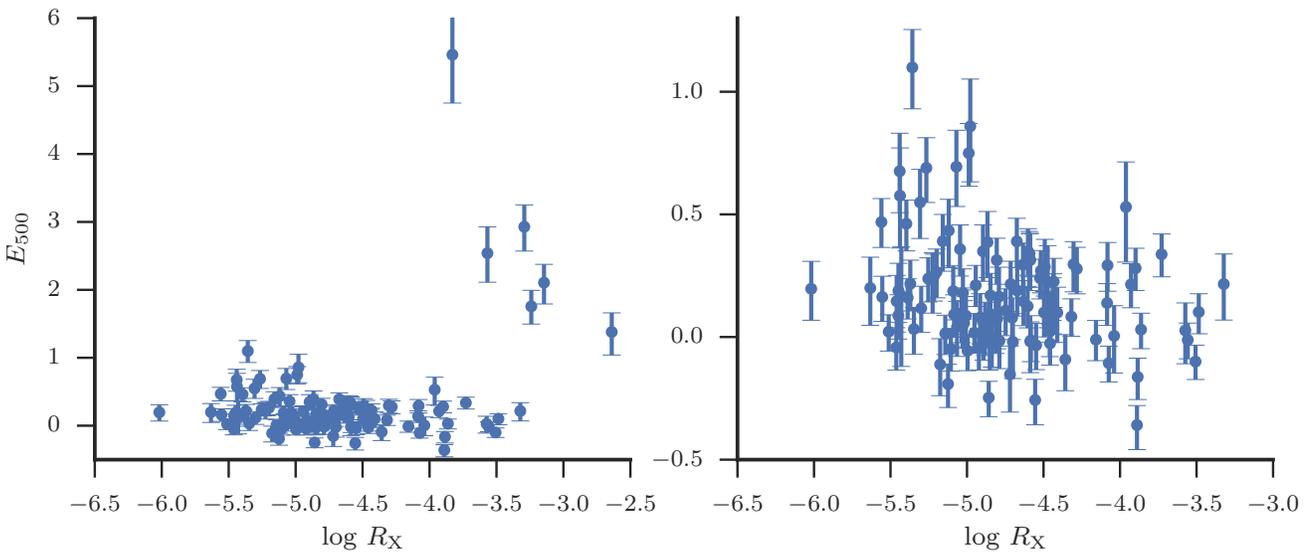}
\caption{\label{fig:excess_vs_rx} The correlation between radio loudness defined as $R_{\rm X} = L_{\rm 1.4\,GHz}/L_{\rm 14-195\, keV}$ and the 500 \um{} excess emission, $E_{\rm 500}$. The left panel displays all sources, while the right panel zooms in on the majority of the BAT AGN with $E_{\rm 500} < 1.0$. Error bars encompass the 68\% confidence region for $E_{\rm 500}$.}
\end{center}
\end{figure*}

Figure~\ref{fig:excess_vs_rx} plots $E_{\rm 500}$ against $R_{\rm X}$ to test our hypothesis that the excess 500 \um{} emission is related to the radio loudness of the AGN. In the left panel we plot all of the sources together to show the full range of $E_{\rm 500}$. Indeed, the six AGN with the largest values of $E_{\rm 500}$ exhibit high values of radio loudness ($\log\,R_{\rm X} > -4.0$). These six AGN are HB 890241+622, 2MASX J23272195+1524375, 3C 111.0, 3C 120, Pictor A, and PKS 2331-240 and all are well known radio-loud AGN. They correspond to the six sources in Figures~\ref{fig:color-color} and \ref{fig:color-color_excess} that lie in the lower left hand corner. Further, the lone HRS galaxy seen in Figure~\ref{fig:color-color} among the six BAT AGN is the radio galaxy M87, whose jets and radio activity have been studied extensively. Based on this, we prescribe color cutoffs that can easily separate radio-loud AGN from radio-quiet AGN and normal star-forming galaxies: $F_{250}/F_{350} < 1.5$ and $F_{350}/F_{500} < 1.5$ (see dashed lines in Figure~\ref{fig:color-color_excess}).

While radio-loudness can explain the most extreme values of $E_{\rm 500}$, it does not explain the more moderate ones. In the right panel of Figure~\ref{fig:excess_vs_rx}, we zoom in on the AGN with $E_{\rm 500} < 1.0$. Visually there does not appear to be any strong correlation between $R_{\rm X}$ and $E_{\rm 500}$ and the Spearman rank correlation coefficient between them is -0.15, weak and in the opposite sense of what would be expected if synchrotron emission was contaminating the 500 \um{} emission.
  
To explore even further, we analyzed the correlations between $E_{\rm 500}$ and two AGN-related indicators, the \swift/BAT luminosity, $L_{14-195\,keV}$, and the 3.4 to 4.6 \um{} flux ratio ($W1/W2$). The 3.4 ($W1$) and 4.6 ($W2$) \um{} fluxes for the BAT AGN were obtained from the \textit{Wide-field Infrared Survey Explorer} \citep[\textit{WISE};][]{Wright:2010fk} AllWISE catalog accessed through the \textit{NASA/IPAC} Infrared Science Archive (IRSA)\footnote{\url{http://irsa.ipac.caltech.edu/Missions/wise.html}}. Details of the compilation of \textit{WISE} fluxes for the BAT AGN will be available in an upcoming publication (Shimizu et al. in preparation).

\citet{Winter:2012yq} showed that $L_{14-195\,keV}$ can be used as a measure of the intrinsic bolometric luminosity of the AGN, unaffected by host galaxy contamination or line-of-sight absorption. $W1/W2$ has been shown to be an effective discriminator between AGN-dominated and normal star-forming galaxies that has both high reliability and completeness \citep{Stern:2012mz}\footnote{\citet{Stern:2012mz} prescribe a cutoff of $W1 - W2 \geq 0.8$  in magnitude units for selecting AGN. In flux units this changes to $W1/W2 \leq 0.86$}. \citet{Stern:2012mz} also show that as the fraction of emission coming from the host galaxy increases $W1/W2$ increases as well making it a good measure of the relative contribution of the AGN to the infrared luminosity.

\begin{figure*}
\begin{center}
\includegraphics[width=\textwidth]{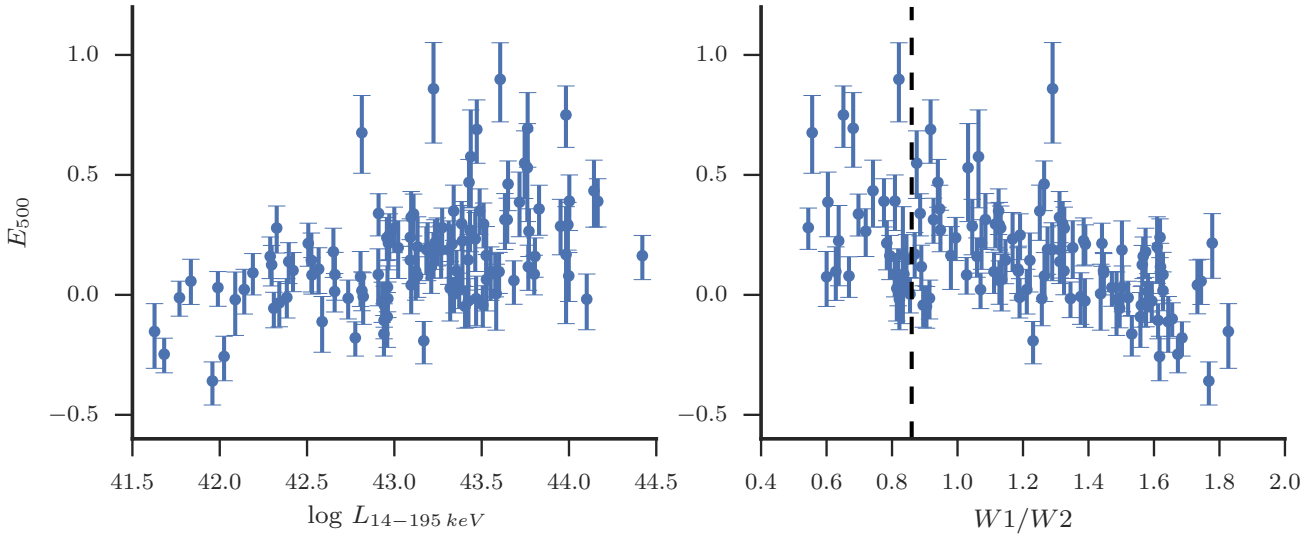}
\caption{\label{fig:excess_vs_agn} Relationships between $E_{\rm 500}$ and the hard X-ray luminosity (left), $L_{\rm 14-195\,keV}$, as a proxy for the intrinsic AGN strength, and $W1/W2$ (right), the ratio of the 3.4 and 4.6 \um{} flux densities, as a proxy for the relative strength of the AGN to host galaxy emission. In the right panel we also plot as a dashed line, the cutoff (0.86) for AGN-dominated galaxies prescribed by \citet{Stern:2012mz} after converting to flux densities. In both panels we removed sources with $E_{\rm 500} > 1.0$ that are associated with radio-loud AGN. Error bars encompass the 68\% confidence region for $E_{\rm 500}$.}
\end{center}
\end{figure*}

Fig.~\ref{fig:excess_vs_agn} shows the relationships between both $L_{\rm 14-195\,keV}$ (left panel) and $W1/W2$ (right panel) with $E_{\rm 500}$ after removing the six radio-loud AGN. Both parameters display noticeable correlations with $E_{\rm 500}$ with $L_{\rm 14-195\,keV}$ positively correlated and $W1/W2$ negatively correlated. We calculated Spearman rank correlation coefficients finding values of 0.49 and -0.49 respectively. Pearson correlation coefficients are 0.30 and -0.50 respectively reflecting the more linear relationship between $E_{\rm 500}$ and $W1/W2$ than the one between $E_{\rm 500}$ and $L_{\rm 14-195\,keV}$. All correlations have a probability of a null correlation less than 0.01\%. In the right panel we also plot the \citet{Stern:2012mz} cutoff for AGN-dominated galaxies where values to the left of this line indicate AGN-dominated colors. 

Both panels indicate that the strength of the AGN in the host galaxy is possibly having an effect on the SPIRE colors. A stronger AGN in relation to the host galaxy is causing deviations from a standard modified blackbody in the form of a small but noticeable 500 \um{} offset.

Without longer wavelength data, however, its impossible to determine the exact cause of the 500 \um{} excess so we can only speculate. Submillimeter excess emission has been observed in a number of objects including dwarf and normal star-forming galaxies \citep[e.g.][]{Galametz:2009cl,Galametz:2011ao,Dale:2012dq,Remy-Ruyer:2013kx} as well as the Small and Large Magellanic Clouds \citep{Bot:2010zm,Gordon:2010ix} and even our own Milky Way \citep{Paradis:2012oj}. Various explanations have been proposed including the presence of a very cold ($T \sim 10$ K) component \citep{Galametz:2009cl,Galametz:2011ao,OHalloran:2010wt}, grain coagulation that causes the emissivity to increase for colder temperatures \citep{Paradis:2009hb}, fluctuations in the Cosmic Microwave Background \citep{Planck-Collaboration:2011uk}, and an increase in magnetic material in the ISM \citep{Draine:2012vf}. While all of these explanations are certainly still possible to explain the excess seen in the BAT AGN, they lack any direct connection to the strength of the AGN. Further, a key result from all of the previous work is that the submillimeter excess is more prevalent in very metal-poor galaxies ($12 + \log(\rm{O/H} \lesssim 8.3$). All of the BAT AGN reside in high stellar mass galaxies \citep{Koss:2011vn} and given the mass-metallicity relationship \citep{Tremonti:2004fq} should also be quite metal rich. 


Rather, we speculate the excess is related to radio emission more closely associated with the AGN itself. Several studies of the radio properties of AGN have revealed a millimeter excess around 100 GHz \citep{Doi:2005wj,Doi:2011si,Behar:2015le,Scharwachter:2015ez} that is likely due to either an inverted or flat SED between cm and mm wavelengths. Because \citet{Doi:2011si} found the excess mainly in low luminosity AGN similar to Sgr A*, they invoked advection dominated accretion flows (ADAF) that produce compact nuclear jets to explain the inverter or flat SEDs. However the sample of \citet{Behar:2015le} was composed of X-ray bright AGN including high Eddington ratio ($L_{\rm bol}/L_{\rm Edd}$, a measure of the accretion rate relative to the Eddington limit) sources where an ADAF is unlikely. \citet{Behar:2015le} instead use the radio-to-X-ray luminosity ratio to argue that the high-frequency radio emission originates near the X-ray corona of the accretion disk given the ratio's similarity to that found for stellar coronal mass ejections \citep[e.g][]{Bastian:1998tx} as well as the compact nature of the radio emission. Magnetic activity around the accretion disk in the core of the AGN would then be responsible for the excess and if magnetic activity increases with $L_{\rm bol}/L_{\rm Edd}$, this could explain the relationship seen with $L_{\rm 14-195\,keV}$ as well as $W1/W2$. This strengthens the need for a more comprehensive survey of AGN in the mm wavelength range as it could clearly reveal interesting physics possibly occurring near the accretion disk.

\section{Conclusions}
We have produced the \herschel/SPIRE maps for 313 AGN selected from the \swift/BAT 58 month catalog in three wavebands: 250, 350, and 500 \um. Combined with the PACS photometry from \citet{Melendez:2014yu}, the SPIRE flux densities presented in this Paper form the complete FIR SEDs for a large, nearby, and relatively unbiased sample of AGN.  We used two methods for measuring the flux densities: timeline fitting for point sources and aperture photometry for extended and undetected sources. We summarize below the results of our statistical analysis and comparison to the \herschel{} Reference Survey sample of normal star-forming galaxies.
\begin{itemize}
\item Sy 2s are detected at a higher rate than Sy 1s, and after accounting for upper limits, Sy 2's have slightly higher SPIRE luminosities than Sy 1's. However the effect is small and indicates that on average, the global FIR properties of AGN are independent of orientation.

\item Using a partial correlation survival analysis to account for the luminosity-distance effect and upper limits, we find all of the \herschel{} luminosities are correlated with each other suggesting the process (or processes) producing the emission from 70--500 \um{} is connected. Luminosities with the smallest wavelength difference (i.e. 160 and 250 \um) are much more correlated than pairs further apart (i.e. 70 and 500 \um), in agreement with different temperature components associated with different wavebands. While this could point to the AGN affecting the shorter wavebands more than the longer ones and increasing the scatter, it can also be explained by an increased contribution from older stellar populations to the emission at longer wavelengths.

\item None of the SPIRE luminosities are well correlated with the 14--195 keV luminosity, a proxy for the bolometric AGN luminosity. The AGN, in general, is unlikely to be strongly affecting either the 250, 350, or 500 \um{} emission, however Sy 1s do show a very weak correlation  at 250 and 350 \um{}. Removing CT sources does not improve the correlation for Sy 2's. It remains to be seen what the exact explanation is for the difference in correlations between Sy 1s and Sy 2s but possible explanations include a direct link between star-formation and AGN luminosity that is evident only at high luminosity or increased contamination by the AGN.

\item We compared the SPIRE colors, $F_{250}/F_{350}$ and $F_{350}/F_{500}$, with the colors of the HRS galaxies. The BAT AGN have statistically similar SPIRE color distributions as the high stellar mass ($\log M_{*} > 9.5$ M$_{\odot}$) HRS galaxies. This further emphasizes that on average, the FIR emission of AGN host galaxies is likely produced by cold dust in the ISM heated by stellar radiation just as in normal star-forming galaxies without an AGN.

\item We did find anomalous colors for 6 BAT AGN with $F_{250}/F_{350} < 1.5$ and $F_{350}/F_{500} < 1.5$. The FIR SEDs for these AGN are dominated by synchrotron emission from a radio jet rather than thermally heated dust.

\item Another group of AGN with less anomalous colors but still removed from the main locus were analyzed by fitting the SEDs with a modified blackbody and calculating a 500 \um{} excess. We found the 500 \um{} excess is not related to radio loudness, but is well correlated with the 14--195 keV luminosity and $W1/W2$ (3.4/4.6 \um) color from \textit{WISE}. We speculate this is possibly related to the millimeter excess emission recently seen in AGN caused by coronal emission above the accretion disk.

\end{itemize}

Future work will focus on combining the photometry from \citet{Melendez:2014yu} and this Paper as well as archival data to perform detailed SED modeling to investigate the local starburst-AGN connection and the AGN contribution to the FIR.

 
\section*{Acknowledgements}
We thank the anonymous referee whose comments and suggestions contributed and improved the quality of this Paper.

The Herschel spacecraft was designed, built, tested, and launched under a contract to ESA managed by the Herschel/Planck Project team by an industrial consortium under the overall responsibility of the prime contractor Thales Alenia Space (Cannes), and including Astrium (Friedrichshafen) responsible for the payload module and for system testing at spacecraft level, Thales Alenia Space (Turin) responsible for the service module, and Astrium (Toulouse) responsible for the telescope, with in excess of a hundred subcontractors. SPIRE has been developed by a consortium of institutes led by Cardiff University (UK) and including Univ. Lethbridge (Canada); NAOC (China); CEA, LAM (France); IFSI, Univ. Padua (Italy); IAC (Spain); Stockholm Observatory (Sweden); Imperial College London, RAL, UCL-MSSL, UKATC, Univ. Sussex (UK); and Caltech, JPL, NHSC, Univ. Colorado (USA). This development has been supported by national funding agencies: CSA (Canada); NAOC (China); CEA, CNES, CNRS (France); ASI (Italy); MCINN (Spain); SNSB (Sweden); STFC, UKSA (UK); and NASA (USA). 

This research has made use of the NASA/IPAC Extragalactic Database (NED), which is operated by the Jet Propulsion Laboratory, California Institute of Technology, under contract with the National Aeronautics and Space Administration.

This publication makes use of data products from the Wide-field Infrared Survey Explorer, which is a joint project of the University of California, Los Angeles, and the Jet Propulsion Laboratory/California Institute of Technology, funded by the National Aeronautics and Space Administration.

All figures in this publication were not possible without the Python package \texttt{MATPLOTLIB} \citep{Hunter:2007}.

\bibliographystyle{mn2e}
\footnotesize{
\bibliography{mn-jour,bibliography/converted_to_latex.bib,bibliography/my_bib.bib}%

\begin{thebibliography}{}
\makeatletter
\relax
\def\mn@urlcharsother{\let\do\@makeother \do\$\do\&\do\#\do\^\do\_\do\%\do\~}
\def\mn@doi{\begingroup\mn@urlcharsother \@ifnextchar [ {\mn@doi@}
  {\mn@doi@[]}}
\def\mn@doi@[#1]#2{\def\@tempa{#1}\ifx\@tempa\@empty \href
  {http://dx.doi.org/#2} {doi:#2}\else \href {http://dx.doi.org/#2} {#1}\fi
  \endgroup}
\def\mn@eprint#1#2{\mn@eprint@#1:#2::\@nil}
\def\mn@eprint@arXiv#1{\href {http://arxiv.org/abs/#1} {{\tt arXiv:#1}}}
\def\mn@eprint@dblp#1{\href {http://dblp.uni-trier.de/rec/bibtex/#1.xml}
  {dblp:#1}}
\def\mn@eprint@#1:#2:#3:#4\@nil{\def\@tempa {#1}\def\@tempb {#2}\def\@tempc
  {#3}\ifx \@tempc \@empty \let \@tempc \@tempb \let \@tempb \@tempa \fi \ifx
  \@tempb \@empty \def\@tempb {arXiv}\fi \@ifundefined
  {mn@eprint@\@tempb}{\@tempb:\@tempc}{\expandafter \expandafter \csname
  mn@eprint@\@tempb\endcsname \expandafter{\@tempc}}}

\bibitem[\protect\citeauthoryear{Akritas \& Siebert}{Akritas \&
  Siebert}{1996}]{Akritas_1996}
Akritas M.~G.,  Siebert J.,  1996, \mn@doi [Monthly Notices of the Royal
  Astronomical Society] {10.1093/mnras/278.4.919}, 278, 919

\bibitem[\protect\citeauthoryear{{Alatalo} et~al.,}{{Alatalo}
  et~al.}{2011}]{Alatalo:2011lk}
{Alatalo} K.,  et~al., 2011, \mn@doi [\apj] {10.1088/0004-637X/735/2/88}, \href
  {http://adsabs.harvard.edu/abs/2011ApJ...735...88A} {735, 88}

\bibitem[\protect\citeauthoryear{{Aniano}, {Draine}, {Gordon}  \&
  {Sandstrom}}{{Aniano} et~al.}{2011}]{Aniano:2011rr}
{Aniano} G.,  {Draine} B.~T.,  {Gordon} K.~D.,   {Sandstrom} K.,  2011, \mn@doi
  [\pasp] {10.1086/662219}, \href
  {http://adsabs.harvard.edu/abs/2011PASP..123.1218A} {123, 1218}

\bibitem[\protect\citeauthoryear{{Antonucci}}{{Antonucci}}{1993}]{Antonucci:1993os}
{Antonucci} R.,  1993, \mn@doi [\araa] {10.1146/annurev.aa.31.090193.002353},
  \href {http://adsabs.harvard.edu/abs/1993ARA%26A..31..473A} {31, 473}

\bibitem[\protect\citeauthoryear{{Astropy Collaboration} et~al.,}{{Astropy
  Collaboration} et~al.}{2013}]{Astropy:2013ek}
{Astropy Collaboration} et~al., 2013, \mn@doi [\aap]
  {10.1051/0004-6361/201322068}, \href
  {http://adsabs.harvard.edu/abs/2013A%26A...558A..33A} {558, A33}

\bibitem[\protect\citeauthoryear{{Baes} et~al.,}{{Baes}
  et~al.}{2010}]{Baes:2010ek}
{Baes} M.,  et~al., 2010, \mn@doi [\aap] {10.1051/0004-6361/201014555}, \href
  {http://adsabs.harvard.edu/abs/2010A%26A...518L..53B} {518, L53}

\bibitem[\protect\citeauthoryear{Barthelmy et~al.,}{Barthelmy
  et~al.}{2005}]{Barthelmy_2005}
Barthelmy S.~D.,  et~al., 2005, \mn@doi [Space Sci Rev]
  {10.1007/s11214-005-5096-3}, 120, 143

\bibitem[\protect\citeauthoryear{{Barvainis}}{{Barvainis}}{1987}]{Barvainis:1987ty}
{Barvainis} R.,  1987, \mn@doi [\apj] {10.1086/165571}, \href
  {http://adsabs.harvard.edu/abs/1987ApJ...320..537B} {320, 537}

\bibitem[\protect\citeauthoryear{{Bastian}, {Benz}  \& {Gary}}{{Bastian}
  et~al.}{1998}]{Bastian:1998tx}
{Bastian} T.~S.,  {Benz} A.~O.,   {Gary} D.~E.,  1998, \mn@doi [\araa]
  {10.1146/annurev.astro.36.1.131}, \href
  {http://adsabs.harvard.edu/abs/1998ARA%26A..36..131B} {36, 131}

\bibitem[\protect\citeauthoryear{{Baumgartner}, {Tueller}, {Markwardt},
  {Skinner}, {Barthelmy}, {Mushotzky}, {Evans}  \& {Gehrels}}{{Baumgartner}
  et~al.}{2012}]{Baumgartner:2012gf}
{Baumgartner} W.~H.,  {Tueller} J.,  {Markwardt} C.~B.,  {Skinner} G.~K.,
  {Barthelmy} S.,  {Mushotzky} R.~F.,  {Evans} P.,   {Gehrels} N.,  2012,
  preprint, \href {http://adsabs.harvard.edu/abs/2012arXiv1212.3336B} {}
  (\mn@eprint {arXiv} {1212.3336})

\bibitem[\protect\citeauthoryear{{Baumgartner}, {Tueller}, {Markwardt},
  {Skinner}, {Barthelmy}, {Mushotzky}, {Evans}  \& {Gehrels}}{{Baumgartner}
  et~al.}{2013}]{Baumgartner:2013fq}
{Baumgartner} W.~H.,  {Tueller} J.,  {Markwardt} C.~B.,  {Skinner} G.~K.,
  {Barthelmy} S.,  {Mushotzky} R.~F.,  {Evans} P.~A.,   {Gehrels} N.,  2013,
  \mn@doi [\apjs] {10.1088/0067-0049/207/2/19}, \href
  {http://adsabs.harvard.edu/abs/2013ApJS..207...19B} {207, 19}

\bibitem[\protect\citeauthoryear{{Becker}, {White}  \& {Edwards}}{{Becker}
  et~al.}{1991}]{Becker:1991qd}
{Becker} R.~H.,  {White} R.~L.,   {Edwards} A.~L.,  1991, \mn@doi [\apjs]
  {10.1086/191529}, \href {http://adsabs.harvard.edu/abs/1991ApJS...75....1B}
  {75, 1}

\bibitem[\protect\citeauthoryear{{Becker}, {White}  \& {Helfand}}{{Becker}
  et~al.}{1995}]{Becker:1995lq}
{Becker} R.~H.,  {White} R.~L.,   {Helfand} D.~J.,  1995, \mn@doi [\apj]
  {10.1086/176166}, \href {http://adsabs.harvard.edu/abs/1995ApJ...450..559B}
  {450, 559}

\bibitem[\protect\citeauthoryear{{Behar}, {Baldi}, {Laor}, {Horesh}, {Stevens}
  \& {Tzioumis}}{{Behar} et~al.}{2015}]{Behar:2015le}
{Behar} E.,  {Baldi} R.~D.,  {Laor} A.,  {Horesh} A.,  {Stevens} J.,
  {Tzioumis} T.,  2015, \mn@doi [\mnras] {10.1093/mnras/stv988}, \href
  {http://adsabs.harvard.edu/abs/2015MNRAS.451.5036B} {451, 5036}

\bibitem[\protect\citeauthoryear{{Bendo} et~al.,}{{Bendo}
  et~al.}{2010}]{Bendo:2010kq}
{Bendo} G.~J.,  et~al., 2010, \mn@doi [\aap] {10.1051/0004-6361/201014568},
  \href {http://adsabs.harvard.edu/abs/2010A%26A...518L..65B} {518, L65}

\bibitem[\protect\citeauthoryear{{Bendo} et~al.,}{{Bendo}
  et~al.}{2012}]{Bendo:2012lr}
{Bendo} G.~J.,  et~al., 2012, \mn@doi [\mnras]
  {10.1111/j.1365-2966.2011.19735.x}, \href
  {http://adsabs.harvard.edu/abs/2012MNRAS.419.1833B} {419, 1833}

\bibitem[\protect\citeauthoryear{{Bendo} et~al.,}{{Bendo}
  et~al.}{2013}]{Bendo:2013sd}
{Bendo} G.~J.,  et~al., 2013, \mn@doi [\mnras] {10.1093/mnras/stt948}, \href
  {http://adsabs.harvard.edu/abs/2013MNRAS.433.3062B} {433, 3062}

\bibitem[\protect\citeauthoryear{{Bendo} et~al.,}{{Bendo}
  et~al.}{2015}]{Bendo:2015lr}
{Bendo} G.~J.,  et~al., 2015, \mn@doi [\mnras] {10.1093/mnras/stu1841}, \href
  {http://adsabs.harvard.edu/abs/2015MNRAS.448..135B} {448, 135}

\bibitem[\protect\citeauthoryear{{Berney} et~al.,}{{Berney}
  et~al.}{2015}]{Berney:2015lr}
{Berney} S.,  et~al., 2015, \mn@doi [\mnras] {10.1093/mnras/stv2181}, \href
  {http://adsabs.harvard.edu/abs/2015MNRAS.454.3622B} {454, 3622}

\bibitem[\protect\citeauthoryear{{Bertin} \& {Arnouts}}{{Bertin} \&
  {Arnouts}}{1996}]{Bertin:1996fk}
{Bertin} E.,  {Arnouts} S.,  1996, \mn@doi [\aaps] {10.1051/aas:1996164}, \href
  {http://adsabs.harvard.edu/abs/1996A%26AS..117..393B} {117, 393}

\bibitem[\protect\citeauthoryear{{Best}}{{Best}}{2007}]{Best:2007vn}
{Best} P.~N.,  2007, \mn@doi [\nar] {10.1016/j.newar.2006.11.014}, \href
  {http://adsabs.harvard.edu/abs/2007NewAR..51..168B} {51, 168}

\bibitem[\protect\citeauthoryear{{Bock}, {Large}  \& {Sadler}}{{Bock}
  et~al.}{1999}]{Bock:1999fp}
{Bock} D.~C.-J.,  {Large} M.~I.,   {Sadler} E.~M.,  1999, \mn@doi [\aj]
  {10.1086/300786}, \href {http://adsabs.harvard.edu/abs/1999AJ....117.1578B}
  {117, 1578}

\bibitem[\protect\citeauthoryear{{Boquien} et~al.,}{{Boquien}
  et~al.}{2011}]{Boquien:2011qf}
{Boquien} M.,  et~al., 2011, \mn@doi [\aj] {10.1088/0004-6256/142/4/111}, \href
  {http://adsabs.harvard.edu/abs/2011AJ....142..111B} {142, 111}

\bibitem[\protect\citeauthoryear{Boselli, Gavazzi  \& Sanvito}{Boselli
  et~al.}{2003}]{Boselli_2003}
Boselli A.,  Gavazzi G.,   Sanvito G.,  2003, \mn@doi [Astronomy and
  Astrophysics] {10.1051/0004-6361:20030219}, 402, 37

\bibitem[\protect\citeauthoryear{{Boselli} et~al.,}{{Boselli}
  et~al.}{2010a}]{Boselli:2010fj}
{Boselli} A.,  et~al., 2010a, \mn@doi [\pasp] {10.1086/651535}, \href
  {http://adsabs.harvard.edu/abs/2010PASP..122..261B} {122, 261}

\bibitem[\protect\citeauthoryear{{Boselli} et~al.,}{{Boselli}
  et~al.}{2010b}]{Boselli:2010fr}
{Boselli} A.,  et~al., 2010b, \mn@doi [\aap] {10.1051/0004-6361/201014534},
  \href {http://adsabs.harvard.edu/abs/2010A%26A...518L..61B} {518, L61}

\bibitem[\protect\citeauthoryear{{Boselli} et~al.,}{{Boselli}
  et~al.}{2012}]{Boselli:2012qv}
{Boselli} A.,  et~al., 2012, \mn@doi [\aap] {10.1051/0004-6361/201118602},
  \href {http://adsabs.harvard.edu/abs/2012A%26A...540A..54B} {540, A54}

\bibitem[\protect\citeauthoryear{{Bot}, {Ysard}, {Paradis}, {Bernard},
  {Lagache}, {Israel}  \& {Wall}}{{Bot} et~al.}{2010}]{Bot:2010zm}
{Bot} C.,  {Ysard} N.,  {Paradis} D.,  {Bernard} J.~P.,  {Lagache} G.,
  {Israel} F.~P.,   {Wall} W.~F.,  2010, \mn@doi [\aap]
  {10.1051/0004-6361/201014986}, \href
  {http://adsabs.harvard.edu/abs/2010A%26A...523A..20B} {523, A20}

\bibitem[\protect\citeauthoryear{{Bower}, {Benson}, {Malbon}, {Helly}, {Frenk},
  {Baugh}, {Cole}  \& {Lacey}}{{Bower} et~al.}{2006}]{Bower:2006gf}
{Bower} R.~G.,  {Benson} A.~J.,  {Malbon} R.,  {Helly} J.~C.,  {Frenk} C.~S.,
  {Baugh} C.~M.,  {Cole} S.,   {Lacey} C.~G.,  2006, \mn@doi [\mnras]
  {10.1111/j.1365-2966.2006.10519.x}, \href
  {http://adsabs.harvard.edu/abs/2006MNRAS.370..645B} {370, 645}

\bibitem[\protect\citeauthoryear{Burgarella et~al.,}{Burgarella
  et~al.}{2013}]{Burgarella_2013}
Burgarella D.,  et~al., 2013, \mn@doi [Astronomy {\&} Astrophysics]
  {10.1051/0004-6361/201321651}, 554, A70

\bibitem[\protect\citeauthoryear{{Burlon}, {Ajello}, {Greiner}, {Comastri},
  {Merloni}  \& {Gehrels}}{{Burlon} et~al.}{2011}]{Burlon:2011pi}
{Burlon} D.,  {Ajello} M.,  {Greiner} J.,  {Comastri} A.,  {Merloni} A.,
  {Gehrels} N.,  2011, \mn@doi [\apj] {10.1088/0004-637X/728/1/58}, \href
  {http://adsabs.harvard.edu/abs/2011ApJ...728...58B} {728, 58}

\bibitem[\protect\citeauthoryear{{Calzetti}, {Armus}, {Bohlin}, {Kinney},
  {Koornneef}  \& {Storchi-Bergmann}}{{Calzetti}
  et~al.}{2000}]{Calzetti:2000fk}
{Calzetti} D.,  {Armus} L.,  {Bohlin} R.~C.,  {Kinney} A.~L.,  {Koornneef} J.,
   {Storchi-Bergmann} T.,  2000, \mn@doi [\apj] {10.1086/308692}, \href
  {http://adsabs.harvard.edu/abs/2000ApJ...533..682C} {533, 682}

\bibitem[\protect\citeauthoryear{{Chen} et~al.,}{{Chen}
  et~al.}{2013}]{Chen:2013uq}
{Chen} C.-T.~J.,  et~al., 2013, \mn@doi [\apj] {10.1088/0004-637X/773/1/3},
  \href {http://adsabs.harvard.edu/abs/2013ApJ...773....3C} {773, 3}

\bibitem[\protect\citeauthoryear{{Cid Fernandes}, {Heckman}, {Schmitt},
  {Gonz{\'a}lez Delgado}  \& {Storchi-Bergmann}}{{Cid Fernandes}
  et~al.}{2001}]{Cid-Fernandes:2001uq}
{Cid Fernandes} R.,  {Heckman} T.,  {Schmitt} H.,  {Gonz{\'a}lez Delgado}
  R.~M.,   {Storchi-Bergmann} T.,  2001, \mn@doi [\apj] {10.1086/322449}, \href
  {http://adsabs.harvard.edu/abs/2001ApJ...558...81C} {558, 81}

\bibitem[\protect\citeauthoryear{{Ciesla} et~al.,}{{Ciesla}
  et~al.}{2012}]{Ciesla:2012lq}
{Ciesla} L.,  et~al., 2012, \mn@doi [\aap] {10.1051/0004-6361/201219216}, \href
  {http://adsabs.harvard.edu/abs/2012A%26A...543A.161C} {543, A161}

\bibitem[\protect\citeauthoryear{{Ciesla} et~al.,}{{Ciesla}
  et~al.}{2015}]{Ciesla:2015qr}
{Ciesla} L.,  et~al., 2015, \mn@doi [\aap] {10.1051/0004-6361/201425252}, \href
  {http://adsabs.harvard.edu/abs/2015A%26A...576A..10C} {576, A10}

\bibitem[\protect\citeauthoryear{{Cirasuolo}, {Magliocchetti}, {Celotti}  \&
  {Danese}}{{Cirasuolo} et~al.}{2003a}]{Cirasuolo:2003zl}
{Cirasuolo} M.,  {Magliocchetti} M.,  {Celotti} A.,   {Danese} L.,  2003a,
  \mn@doi [\mnras] {10.1046/j.1365-8711.2003.06485.x}, \href
  {http://adsabs.harvard.edu/abs/2003MNRAS.341..993C} {341, 993}

\bibitem[\protect\citeauthoryear{{Cirasuolo}, {Celotti}, {Magliocchetti}  \&
  {Danese}}{{Cirasuolo} et~al.}{2003b}]{Cirasuolo:2003rm}
{Cirasuolo} M.,  {Celotti} A.,  {Magliocchetti} M.,   {Danese} L.,  2003b,
  \mn@doi [\mnras] {10.1046/j.1365-2966.2003.07105.x}, \href
  {http://adsabs.harvard.edu/abs/2003MNRAS.346..447C} {346, 447}

\bibitem[\protect\citeauthoryear{{Condon}, {Cotton}, {Greisen}, {Yin},
  {Perley}, {Taylor}  \& {Broderick}}{{Condon} et~al.}{1998}]{Condon:1998eu}
{Condon} J.~J.,  {Cotton} W.~D.,  {Greisen} E.~W.,  {Yin} Q.~F.,  {Perley}
  R.~A.,  {Taylor} G.~B.,   {Broderick} J.~J.,  1998, \mn@doi [\aj]
  {10.1086/300337}, \href {http://adsabs.harvard.edu/abs/1998AJ....115.1693C}
  {115, 1693}

\bibitem[\protect\citeauthoryear{{Cortese} et~al.,}{{Cortese}
  et~al.}{2012}]{Cortese:2012fj}
{Cortese} L.,  et~al., 2012, \mn@doi [\aap] {10.1051/0004-6361/201219312},
  \href {http://adsabs.harvard.edu/abs/2012A%26A...544A.101C} {544, A101}

\bibitem[\protect\citeauthoryear{{Cortese} et~al.,}{{Cortese}
  et~al.}{2014}]{Cortese:2014qq}
{Cortese} L.,  et~al., 2014, \mn@doi [\mnras] {10.1093/mnras/stu175}, \href
  {http://adsabs.harvard.edu/abs/2014MNRAS.440..942C} {440, 942}

\bibitem[\protect\citeauthoryear{{Cowie}, {Barger}, {Bautz}, {Brandt}  \&
  {Garmire}}{{Cowie} et~al.}{2003}]{Cowie:2003fk}
{Cowie} L.~L.,  {Barger} A.~J.,  {Bautz} M.~W.,  {Brandt} W.~N.,   {Garmire}
  G.~P.,  2003, \mn@doi [\apjl] {10.1086/368404}, \href
  {http://adsabs.harvard.edu/abs/2003ApJ...584L..57C} {584, L57}

\bibitem[\protect\citeauthoryear{{Croton} et~al.,}{{Croton}
  et~al.}{2006}]{Croton:2006kx}
{Croton} D.~J.,  et~al., 2006, \mn@doi [\mnras]
  {10.1111/j.1365-2966.2005.09675.x}, \href
  {http://adsabs.harvard.edu/abs/2006MNRAS.365...11C} {365, 11}

\bibitem[\protect\citeauthoryear{{Dale} \& {Helou}}{{Dale} \&
  {Helou}}{2002}]{Dale:2002ty}
{Dale} D.~A.,  {Helou} G.,  2002, \mn@doi [\apj] {10.1086/341632}, \href
  {http://adsabs.harvard.edu/abs/2002ApJ...576..159D} {576, 159}

\bibitem[\protect\citeauthoryear{{Dale} et~al.,}{{Dale}
  et~al.}{2007}]{Dale:2007fk}
{Dale} D.~A.,  et~al., 2007, \mn@doi [\apj] {10.1086/510362}, \href
  {http://adsabs.harvard.edu/abs/2007ApJ...655..863D} {655, 863}

\bibitem[\protect\citeauthoryear{{Dale} et~al.,}{{Dale}
  et~al.}{2012}]{Dale:2012dq}
{Dale} D.~A.,  et~al., 2012, \mn@doi [\apj] {10.1088/0004-637X/745/1/95}, \href
  {http://adsabs.harvard.edu/abs/2012ApJ...745...95D} {745, 95}

\bibitem[\protect\citeauthoryear{{Diamond-Stanic} \& {Rieke}}{{Diamond-Stanic}
  \& {Rieke}}{2012}]{Diamond-Stanic:2012rw}
{Diamond-Stanic} A.~M.,  {Rieke} G.~H.,  2012, \mn@doi [\apj]
  {10.1088/0004-637X/746/2/168}, \href
  {http://adsabs.harvard.edu/abs/2012ApJ...746..168D} {746, 168}

\bibitem[\protect\citeauthoryear{{Dixon} \& {Joseph}}{{Dixon} \&
  {Joseph}}{2011}]{Dixon:2011yq}
{Dixon} T.~G.,  {Joseph} R.~D.,  2011, \mn@doi [\apj]
  {10.1088/0004-637X/740/2/99}, \href
  {http://adsabs.harvard.edu/abs/2011ApJ...740...99D} {740, 99}

\bibitem[\protect\citeauthoryear{{Doi}, {Kameno}  \& {Inoue}}{{Doi}
  et~al.}{2005}]{Doi:2005wj}
{Doi} A.,  {Kameno} S.,   {Inoue} M.,  2005, \mn@doi [\mnras]
  {10.1111/j.1365-2966.2005.09014.x}, \href
  {http://adsabs.harvard.edu/abs/2005MNRAS.360..119D} {360, 119}

\bibitem[\protect\citeauthoryear{{Doi}, {Nakanishi}, {Nagai}, {Kohno}  \&
  {Kameno}}{{Doi} et~al.}{2011}]{Doi:2011si}
{Doi} A.,  {Nakanishi} K.,  {Nagai} H.,  {Kohno} K.,   {Kameno} S.,  2011,
  \mn@doi [\aj] {10.1088/0004-6256/142/5/167}, \href
  {http://adsabs.harvard.edu/abs/2011AJ....142..167D} {142, 167}

\bibitem[\protect\citeauthoryear{{Draine}}{{Draine}}{2003}]{Draine:2003gd}
{Draine} B.~T.,  2003, \mn@doi [\araa]
  {10.1146/annurev.astro.41.011802.094840}, \href
  {http://adsabs.harvard.edu/abs/2003ARA%26A..41..241D} {41, 241}

\bibitem[\protect\citeauthoryear{{Draine} \& {Hensley}}{{Draine} \&
  {Hensley}}{2012}]{Draine:2012vf}
{Draine} B.~T.,  {Hensley} B.,  2012, \mn@doi [\apj]
  {10.1088/0004-637X/757/1/103}, \href
  {http://adsabs.harvard.edu/abs/2012ApJ...757..103D} {757, 103}

\bibitem[\protect\citeauthoryear{{Draine} et~al.,}{{Draine}
  et~al.}{2007}]{Draine:2007rm}
{Draine} B.~T.,  et~al., 2007, \mn@doi [\apj] {10.1086/518306}, \href
  {http://adsabs.harvard.edu/abs/2007ApJ...663..866D} {663, 866}

\bibitem[\protect\citeauthoryear{{Efstathiou} \& {Rowan-Robinson}}{{Efstathiou}
  \& {Rowan-Robinson}}{1995}]{Efstathiou:1995rz}
{Efstathiou} A.,  {Rowan-Robinson} M.,  1995, \mnras, \href
  {http://adsabs.harvard.edu/abs/1995MNRAS.273..649E} {273, 649}

\bibitem[\protect\citeauthoryear{{Elvis} et~al.,}{{Elvis}
  et~al.}{1994}]{Elvis:1994uq}
{Elvis} M.,  et~al., 1994, \mn@doi [\apjs] {10.1086/192093}, \href
  {http://adsabs.harvard.edu/abs/1994ApJS...95....1E} {95, 1}

\bibitem[\protect\citeauthoryear{{Esquej} et~al.,}{{Esquej}
  et~al.}{2014}]{Esquej:2014vl}
{Esquej} P.,  et~al., 2014, \mn@doi [\apj] {10.1088/0004-637X/780/1/86}, \href
  {http://adsabs.harvard.edu/abs/2014ApJ...780...86E} {780, 86}

\bibitem[\protect\citeauthoryear{{Fabian}, {Sanders}, {Allen}, {Crawford},
  {Iwasawa}, {Johnstone}, {Schmidt}  \& {Taylor}}{{Fabian}
  et~al.}{2003}]{Fabian:2003ek}
{Fabian} A.~C.,  {Sanders} J.~S.,  {Allen} S.~W.,  {Crawford} C.~S.,  {Iwasawa}
  K.,  {Johnstone} R.~M.,  {Schmidt} R.~W.,   {Taylor} G.~B.,  2003, \mn@doi
  [\mnras] {10.1046/j.1365-8711.2003.06902.x}, \href
  {http://adsabs.harvard.edu/abs/2003MNRAS.344L..43F} {344, L43}

\bibitem[\protect\citeauthoryear{{Feigelson} \& {Nelson}}{{Feigelson} \&
  {Nelson}}{1985}]{Feigelson:1985lr}
{Feigelson} E.~D.,  {Nelson} P.~I.,  1985, \mn@doi [\apj] {10.1086/163225},
  \href {http://adsabs.harvard.edu/abs/1985ApJ...293..192F} {293, 192}

\bibitem[\protect\citeauthoryear{{Ferrarese} \& {Merritt}}{{Ferrarese} \&
  {Merritt}}{2000}]{Ferrarese:2000gf}
{Ferrarese} L.,  {Merritt} D.,  2000, \mn@doi [\apjl] {10.1086/312838}, \href
  {http://adsabs.harvard.edu/abs/2000ApJ...539L...9F} {539, L9}

\bibitem[\protect\citeauthoryear{Foreman-Mackey, Hogg, Lang  \&
  Goodman}{Foreman-Mackey et~al.}{2013}]{Foreman_Mackey_2013}
Foreman-Mackey D.,  Hogg D.~W.,  Lang D.,   Goodman J.,  2013, \mn@doi
  [Publications of the Astronomical Society of the Pacific] {10.1086/670067},
  125, 306

\bibitem[\protect\citeauthoryear{{Fritz}, {Franceschini}  \&
  {Hatziminaoglou}}{{Fritz} et~al.}{2006}]{Fritz:2006yq}
{Fritz} J.,  {Franceschini} A.,   {Hatziminaoglou} E.,  2006, \mn@doi [\mnras]
  {10.1111/j.1365-2966.2006.09866.x}, \href
  {http://adsabs.harvard.edu/abs/2006MNRAS.366..767F} {366, 767}

\bibitem[\protect\citeauthoryear{{Galametz} et~al.,}{{Galametz}
  et~al.}{2009}]{Galametz:2009cl}
{Galametz} M.,  et~al., 2009, \mn@doi [\aap] {10.1051/0004-6361/200912963},
  \href {http://adsabs.harvard.edu/abs/2009A%26A...508..645G} {508, 645}

\bibitem[\protect\citeauthoryear{{Galametz}, {Madden}, {Galliano}, {Hony},
  {Bendo}  \& {Sauvage}}{{Galametz} et~al.}{2011}]{Galametz:2011ao}
{Galametz} M.,  {Madden} S.~C.,  {Galliano} F.,  {Hony} S.,  {Bendo} G.~J.,
  {Sauvage} M.,  2011, \mn@doi [\aap] {10.1051/0004-6361/201014904}, \href
  {http://adsabs.harvard.edu/abs/2011A%26A...532A..56G} {532, A56}

\bibitem[\protect\citeauthoryear{{Galametz} et~al.,}{{Galametz}
  et~al.}{2012}]{Galametz:2012uq}
{Galametz} M.,  et~al., 2012, \mn@doi [\mnras]
  {10.1111/j.1365-2966.2012.21667.x}, \href
  {http://adsabs.harvard.edu/abs/2012MNRAS.425..763G} {425, 763}

\bibitem[\protect\citeauthoryear{{Gandhi}, {Horst}, {Smette}, {H{\"o}nig},
  {Comastri}, {Gilli}, {Vignali}  \& {Duschl}}{{Gandhi}
  et~al.}{2009}]{Gandhi:2009kx}
{Gandhi} P.,  {Horst} H.,  {Smette} A.,  {H{\"o}nig} S.,  {Comastri} A.,
  {Gilli} R.,  {Vignali} C.,   {Duschl} W.,  2009, \mn@doi [\aap]
  {10.1051/0004-6361/200811368}, \href
  {http://adsabs.harvard.edu/abs/2009A%26A...502..457G} {502, 457}

\bibitem[\protect\citeauthoryear{Gehrels et~al.,}{Gehrels
  et~al.}{2004}]{Gehrels_2004}
Gehrels N.,  et~al., 2004, \mn@doi [{ApJ}] {10.1086/422091}, 611, 1005

\bibitem[\protect\citeauthoryear{Goodman \& Weare}{Goodman \&
  Weare}{2010}]{Goodman_2010}
Goodman J.,  Weare J.,  2010, \mn@doi [Communications in Applied Mathematics
  and Computational Science] {10.2140/camcos.2010.5.65}, 5, 65

\bibitem[\protect\citeauthoryear{{Gordon} et~al.,}{{Gordon}
  et~al.}{2010}]{Gordon:2010ix}
{Gordon} K.~D.,  et~al., 2010, \mn@doi [\aap] {10.1051/0004-6361/201014541},
  \href {http://adsabs.harvard.edu/abs/2010A%26A...518L..89G} {518, L89}

\bibitem[\protect\citeauthoryear{{Griffin} et~al.,}{{Griffin}
  et~al.}{2010}]{Griffin:2010sf}
{Griffin} M.~J.,  et~al., 2010, \mn@doi [\aap] {10.1051/0004-6361/201014519},
  \href {http://adsabs.harvard.edu/abs/2010A%26A...518L...3G} {518, L3}

\bibitem[\protect\citeauthoryear{{Griffith} \& {Wright}}{{Griffith} \&
  {Wright}}{1993}]{Griffith:1993qr}
{Griffith} M.~R.,  {Wright} A.~E.,  1993, \mn@doi [\aj] {10.1086/116545}, \href
  {http://adsabs.harvard.edu/abs/1993AJ....105.1666G} {105, 1666}

\bibitem[\protect\citeauthoryear{{G{\"u}ltekin} et~al.,}{{G{\"u}ltekin}
  et~al.}{2009}]{Gultekin:2009ul}
{G{\"u}ltekin} K.,  et~al., 2009, \mn@doi [\apj] {10.1088/0004-637X/698/1/198},
  \href {http://adsabs.harvard.edu/abs/2009ApJ...698..198G} {698, 198}

\bibitem[\protect\citeauthoryear{{H{\"a}ring} \& {Rix}}{{H{\"a}ring} \&
  {Rix}}{2004}]{Haring:2004ly}
{H{\"a}ring} N.,  {Rix} H.-W.,  2004, \mn@doi [\apjl] {10.1086/383567}, \href
  {http://adsabs.harvard.edu/abs/2004ApJ...604L..89H} {604, L89}

\bibitem[\protect\citeauthoryear{{Harrison}, {Alexander}, {Mullaney}  \&
  {Swinbank}}{{Harrison} et~al.}{2014}]{Harrison:2014xe}
{Harrison} C.~M.,  {Alexander} D.~M.,  {Mullaney} J.~R.,   {Swinbank} A.~M.,
  2014, \mn@doi [\mnras] {10.1093/mnras/stu515}, \href
  {http://adsabs.harvard.edu/abs/2014MNRAS.441.3306H} {441, 3306}

\bibitem[\protect\citeauthoryear{{Hasinger}, {Miyaji}  \& {Schmidt}}{{Hasinger}
  et~al.}{2005}]{Hasinger:2005qy}
{Hasinger} G.,  {Miyaji} T.,   {Schmidt} M.,  2005, \mn@doi [\aap]
  {10.1051/0004-6361:20042134}, \href
  {http://adsabs.harvard.edu/abs/2005A%26A...441..417H} {441, 417}

\bibitem[\protect\citeauthoryear{Hauser \& Dwek}{Hauser \&
  Dwek}{2001}]{Hauser_2001}
Hauser M.~G.,  Dwek E.,  2001, \mn@doi [Annual Review of Astronomy and
  Astrophysics] {10.1146/annurev.astro.39.1.249}, 39, 249

\bibitem[\protect\citeauthoryear{{Hickox}, {Mullaney}, {Alexander}, {Chen},
  {Civano}, {Goulding}  \& {Hainline}}{{Hickox} et~al.}{2014}]{Hickox:2014yq}
{Hickox} R.~C.,  {Mullaney} J.~R.,  {Alexander} D.~M.,  {Chen} C.-T.~J.,
  {Civano} F.~M.,  {Goulding} A.~D.,   {Hainline} K.~N.,  2014, \mn@doi [\apj]
  {10.1088/0004-637X/782/1/9}, \href
  {http://adsabs.harvard.edu/abs/2014ApJ...782....9H} {782, 9}

\bibitem[\protect\citeauthoryear{{Ho} \& {Peng}}{{Ho} \&
  {Peng}}{2001}]{Ho:2001hl}
{Ho} L.~C.,  {Peng} C.~Y.,  2001, \mn@doi [\apj] {10.1086/321524}, \href
  {http://adsabs.harvard.edu/abs/2001ApJ...555..650H} {555, 650}

\bibitem[\protect\citeauthoryear{Hunter}{Hunter}{2007}]{Hunter:2007}
Hunter J.~D.,  2007, Computing In Science \& Engineering, 9, 90

\bibitem[\protect\citeauthoryear{{Ichikawa}, {Ueda}, {Terashima}, {Oyabu},
  {Gandhi}, {Matsuta}  \& {Nakagawa}}{{Ichikawa}
  et~al.}{2012}]{Ichikawa:2012ul}
{Ichikawa} K.,  {Ueda} Y.,  {Terashima} Y.,  {Oyabu} S.,  {Gandhi} P.,
  {Matsuta} K.,   {Nakagawa} T.,  2012, \mn@doi [\apj]
  {10.1088/0004-637X/754/1/45}, \href
  {http://adsabs.harvard.edu/abs/2012ApJ...754...45I} {754, 45}

\bibitem[\protect\citeauthoryear{{Isobe}, {Feigelson}, {Akritas}  \&
  {Babu}}{{Isobe} et~al.}{1990}]{Isobe:1990fk}
{Isobe} T.,  {Feigelson} E.~D.,  {Akritas} M.~G.,   {Babu} G.~J.,  1990,
  \mn@doi [\apj] {10.1086/169390}, \href
  {http://adsabs.harvard.edu/abs/1990ApJ...364..104I} {364, 104}

\bibitem[\protect\citeauthoryear{{Kellermann}, {Sramek}, {Schmidt}, {Shaffer}
  \& {Green}}{{Kellermann} et~al.}{1989}]{Kellermann:1989sf}
{Kellermann} K.~I.,  {Sramek} R.,  {Schmidt} M.,  {Shaffer} D.~B.,   {Green}
  R.,  1989, \mn@doi [\aj] {10.1086/115207}, \href
  {http://adsabs.harvard.edu/abs/1989AJ.....98.1195K} {98, 1195}

\bibitem[\protect\citeauthoryear{{Kennicutt} \& {Evans}}{{Kennicutt} \&
  {Evans}}{2012}]{Kennicutt:2012it}
{Kennicutt} R.~C.,  {Evans} N.~J.,  2012, \mn@doi [\araa]
  {10.1146/annurev-astro-081811-125610}, \href
  {http://adsabs.harvard.edu/abs/2012ARA%26A..50..531K} {50, 531}

\bibitem[\protect\citeauthoryear{{Kessler} et~al.,}{{Kessler}
  et~al.}{1996}]{Kessler:1996wd}
{Kessler} M.~F.,  et~al., 1996, \aap, \href
  {http://adsabs.harvard.edu/abs/1996A%26A...315L..27K} {315, L27}

\bibitem[\protect\citeauthoryear{{Kormendy} \& {Ho}}{{Kormendy} \&
  {Ho}}{2013}]{Kormendy:2013fj}
{Kormendy} J.,  {Ho} L.~C.,  2013, \mn@doi [\araa]
  {10.1146/annurev-astro-082708-101811}, \href
  {http://adsabs.harvard.edu/abs/2013ARA%26A..51..511K} {51, 511}

\bibitem[\protect\citeauthoryear{{Kormendy} \& {Richstone}}{{Kormendy} \&
  {Richstone}}{1995}]{Kormendy:1995mz}
{Kormendy} J.,  {Richstone} D.,  1995, \mn@doi [\araa]
  {10.1146/annurev.aa.33.090195.003053}, \href
  {http://adsabs.harvard.edu/abs/1995ARA%26A..33..581K} {33, 581}

\bibitem[\protect\citeauthoryear{{Koss}, {Mushotzky}, {Veilleux}  \&
  {Winter}}{{Koss} et~al.}{2010}]{Koss:2010nr}
{Koss} M.,  {Mushotzky} R.,  {Veilleux} S.,   {Winter} L.,  2010, \mn@doi
  [\apjl] {10.1088/2041-8205/716/2/L125}, \href
  {http://adsabs.harvard.edu/abs/2010ApJ...716L.125K} {716, L125}

\bibitem[\protect\citeauthoryear{{Koss}, {Mushotzky}, {Veilleux}, {Winter},
  {Baumgartner}, {Tueller}, {Gehrels}  \& {Valencic}}{{Koss}
  et~al.}{2011}]{Koss:2011vn}
{Koss} M.,  {Mushotzky} R.,  {Veilleux} S.,  {Winter} L.~M.,  {Baumgartner} W.,
   {Tueller} J.,  {Gehrels} N.,   {Valencic} L.,  2011, \mn@doi [\apj]
  {10.1088/0004-637X/739/2/57}, \href
  {http://adsabs.harvard.edu/abs/2011ApJ...739...57K} {739, 57}

\bibitem[\protect\citeauthoryear{{LaMassa}, {Heckman}, {Ptak}  \&
  {Urry}}{{LaMassa} et~al.}{2013}]{LaMassa:2013hb}
{LaMassa} S.~M.,  {Heckman} T.~M.,  {Ptak} A.,   {Urry} C.~M.,  2013, \mn@doi
  [\apjl] {10.1088/2041-8205/765/2/L33}, \href
  {http://adsabs.harvard.edu/abs/2013ApJ...765L..33L} {765, L33}

\bibitem[\protect\citeauthoryear{{Lanz}, {Ogle}, {Evans}, {Appleton},
  {Guillard}  \& {Emonts}}{{Lanz} et~al.}{2015}]{Lanz:2015bq}
{Lanz} L.,  {Ogle} P.~M.,  {Evans} D.,  {Appleton} P.~N.,  {Guillard} P.,
  {Emonts} B.,  2015, \mn@doi [\apj] {10.1088/0004-637X/801/1/17}, \href
  {http://adsabs.harvard.edu/abs/2015ApJ...801...17L} {801, 17}

\bibitem[\protect\citeauthoryear{{Laor}}{{Laor}}{2003}]{Laor:2003yg}
{Laor} A.,  2003, ArXiv Astrophysics e-prints, \href
  {http://adsabs.harvard.edu/abs/2003astro.ph.12417L} {}

\bibitem[\protect\citeauthoryear{{Marconi} \& {Hunt}}{{Marconi} \&
  {Hunt}}{2003}]{Marconi:2003ve}
{Marconi} A.,  {Hunt} L.~K.,  2003, \mn@doi [\apjl] {10.1086/375804}, \href
  {http://adsabs.harvard.edu/abs/2003ApJ...589L..21M} {589, L21}

\bibitem[\protect\citeauthoryear{{Matsuta} et~al.,}{{Matsuta}
  et~al.}{2012}]{Matsuta:2012gf}
{Matsuta} K.,  et~al., 2012, \mn@doi [\apj] {10.1088/0004-637X/753/2/104},
  \href {http://adsabs.harvard.edu/abs/2012ApJ...753..104M} {753, 104}

\bibitem[\protect\citeauthoryear{{Mel{\'e}ndez}, {Kraemer}, {Schmitt},
  {Crenshaw}, {Deo}, {Mushotzky}  \& {Bruhweiler}}{{Mel{\'e}ndez}
  et~al.}{2008}]{Melendez:2008pd}
{Mel{\'e}ndez} M.,  {Kraemer} S.~B.,  {Schmitt} H.~R.,  {Crenshaw} D.~M.,
  {Deo} R.~P.,  {Mushotzky} R.~F.,   {Bruhweiler} F.~C.,  2008, \mn@doi [\apj]
  {10.1086/592724}, \href {http://adsabs.harvard.edu/abs/2008ApJ...689...95M}
  {689, 95}

\bibitem[\protect\citeauthoryear{{Mel{\'e}ndez}, {Mushotzky}, {Shimizu},
  {Barger}  \& {Cowie}}{{Mel{\'e}ndez} et~al.}{2014}]{Melendez:2014yu}
{Mel{\'e}ndez} M.,  {Mushotzky} R.~F.,  {Shimizu} T.~T.,  {Barger} A.~J.,
  {Cowie} L.~L.,  2014, \mn@doi [\apj] {10.1088/0004-637X/794/2/152}, \href
  {http://adsabs.harvard.edu/abs/2014ApJ...794..152M} {794, 152}

\bibitem[\protect\citeauthoryear{{Mor} \& {Netzer}}{{Mor} \&
  {Netzer}}{2012}]{Mor:2012fj}
{Mor} R.,  {Netzer} H.,  2012, \mn@doi [\mnras]
  {10.1111/j.1365-2966.2011.20060.x}, \href
  {http://adsabs.harvard.edu/abs/2012MNRAS.420..526M} {420, 526}

\bibitem[\protect\citeauthoryear{{Mullaney}, {Alexander}, {Goulding}  \&
  {Hickox}}{{Mullaney} et~al.}{2011}]{Mullaney:2011yq}
{Mullaney} J.~R.,  {Alexander} D.~M.,  {Goulding} A.~D.,   {Hickox} R.~C.,
  2011, \mn@doi [\mnras] {10.1111/j.1365-2966.2011.18448.x}, \href
  {http://adsabs.harvard.edu/abs/2011MNRAS.414.1082M} {414, 1082}

\bibitem[\protect\citeauthoryear{{Mushotzky}, {Shimizu}, {Mel{\'e}ndez}  \&
  {Koss}}{{Mushotzky} et~al.}{2014}]{Mushotzky:2014ad}
{Mushotzky} R.~F.,  {Shimizu} T.~T.,  {Mel{\'e}ndez} M.,   {Koss} M.,  2014,
  \mn@doi [\apjl] {10.1088/2041-8205/781/2/L34}, \href
  {http://adsabs.harvard.edu/abs/2014ApJ...781L..34M} {781, L34}

\bibitem[\protect\citeauthoryear{{Nenkova}, {Ivezi{\'c}}  \&
  {Elitzur}}{{Nenkova} et~al.}{2002}]{Nenkova:2002ys}
{Nenkova} M.,  {Ivezi{\'c}} {\v Z}.,   {Elitzur} M.,  2002, \mn@doi [\apjl]
  {10.1086/340857}, \href {http://adsabs.harvard.edu/abs/2002ApJ...570L...9N}
  {570, L9}

\bibitem[\protect\citeauthoryear{{Netzer} et~al.,}{{Netzer}
  et~al.}{2007}]{Netzer:2007ve}
{Netzer} H.,  et~al., 2007, \mn@doi [\apj] {10.1086/520716}, \href
  {http://adsabs.harvard.edu/abs/2007ApJ...666..806N} {666, 806}

\bibitem[\protect\citeauthoryear{{Neugebauer} et~al.,}{{Neugebauer}
  et~al.}{1984}]{Neugebauer:1984fp}
{Neugebauer} G.,  et~al., 1984, \mn@doi [\apjl] {10.1086/184209}, \href
  {http://adsabs.harvard.edu/abs/1984ApJ...278L...1N} {278, L1}

\bibitem[\protect\citeauthoryear{{O'Halloran} et~al.,}{{O'Halloran}
  et~al.}{2010}]{OHalloran:2010wt}
{O'Halloran} B.,  et~al., 2010, \mn@doi [\aap] {10.1051/0004-6361/201014580},
  \href {http://adsabs.harvard.edu/abs/2010A%26A...518L..58O} {518, L58}

\bibitem[\protect\citeauthoryear{{Osterbrock}}{{Osterbrock}}{1977}]{Osterbrock:1977fj}
{Osterbrock} D.~E.,  1977, \mn@doi [\apj] {10.1086/155407}, \href
  {http://adsabs.harvard.edu/abs/1977ApJ...215..733O} {215, 733}

\bibitem[\protect\citeauthoryear{{Osterbrock}}{{Osterbrock}}{1981}]{Osterbrock:1981uq}
{Osterbrock} D.~E.,  1981, \mn@doi [\apj] {10.1086/159306}, \href
  {http://adsabs.harvard.edu/abs/1981ApJ...249..462O} {249, 462}

\bibitem[\protect\citeauthoryear{{Ott}}{{Ott}}{2010}]{Ott:2010rm}
{Ott} S.,  2010, in {Mizumoto} Y.,  {Morita} K.-I.,   {Ohishi} M.,  eds,
  Astronomical Society of the Pacific Conference Series Vol. 434, Astronomical
  Data Analysis Software and Systems XIX. p.~139 (\mn@eprint {arXiv}
  {1011.1209})

\bibitem[\protect\citeauthoryear{{Papadopoulos} \& {Allen}}{{Papadopoulos} \&
  {Allen}}{2000}]{Papadopoulos:2000fk}
{Papadopoulos} P.~P.,  {Allen} M.~L.,  2000, \mn@doi [\apj] {10.1086/309066},
  \href {http://adsabs.harvard.edu/abs/2000ApJ...537..631P} {537, 631}

\bibitem[\protect\citeauthoryear{{Papadopoulos} \& {Seaquist}}{{Papadopoulos}
  \& {Seaquist}}{1999}]{Papadopoulos:1999lr}
{Papadopoulos} P.~P.,  {Seaquist} E.~R.,  1999, \mn@doi [\apjl]
  {10.1086/311953}, \href {http://adsabs.harvard.edu/abs/1999ApJ...514L..95P}
  {514, L95}

\bibitem[\protect\citeauthoryear{{Paradis}, {Bernard}  \& {M{\'e}ny}}{{Paradis}
  et~al.}{2009}]{Paradis:2009hb}
{Paradis} D.,  {Bernard} J.-P.,   {M{\'e}ny} C.,  2009, \mn@doi [\aap]
  {10.1051/0004-6361/200811246}, \href
  {http://adsabs.harvard.edu/abs/2009A%26A...506..745P} {506, 745}

\bibitem[\protect\citeauthoryear{{Paradis} et~al.,}{{Paradis}
  et~al.}{2012}]{Paradis:2012oj}
{Paradis} D.,  et~al., 2012, \mn@doi [\aap] {10.1051/0004-6361/201117956},
  \href {http://adsabs.harvard.edu/abs/2012A%26A...537A.113P} {537, A113}

\bibitem[\protect\citeauthoryear{{Petric}, {Ho}, {Flagey}  \&
  {Scoville}}{{Petric} et~al.}{2015}]{Petric:2015fk}
{Petric} A.~O.,  {Ho} L.~C.,  {Flagey} N.~J.~M.,   {Scoville} N.~Z.,  2015,
  \mn@doi [\apjs] {10.1088/0067-0049/219/2/22}, \href
  {http://adsabs.harvard.edu/abs/2015ApJS..219...22P} {219, 22}

\bibitem[\protect\citeauthoryear{{Pier} \& {Krolik}}{{Pier} \&
  {Krolik}}{1992}]{Pier:1992sf}
{Pier} E.~A.,  {Krolik} J.~H.,  1992, \mn@doi [\apj] {10.1086/172042}, \href
  {http://adsabs.harvard.edu/abs/1992ApJ...401...99P} {401, 99}

\bibitem[\protect\citeauthoryear{{Pilbratt} et~al.,}{{Pilbratt}
  et~al.}{2010}]{Pilbratt:2010rz}
{Pilbratt} G.~L.,  et~al., 2010, \mn@doi [\aap] {10.1051/0004-6361/201014759},
  \href {http://adsabs.harvard.edu/abs/2010A%26A...518L...1P} {518, L1}

\bibitem[\protect\citeauthoryear{{Planck Collaboration} et~al.,}{{Planck
  Collaboration} et~al.}{2011}]{Planck-Collaboration:2011uk}
{Planck Collaboration} et~al., 2011, \mn@doi [\aap]
  {10.1051/0004-6361/201116473}, \href
  {http://adsabs.harvard.edu/abs/2011A%26A...536A..17P} {536, A17}

\bibitem[\protect\citeauthoryear{{Planck Collaboration} et~al.,}{{Planck
  Collaboration} et~al.}{2013}]{Planck-Collaboration:2013rt}
{Planck Collaboration} et~al., 2013, preprint, \href
  {http://adsabs.harvard.edu/abs/2013arXiv1303.5088P} {} (\mn@eprint {arXiv}
  {1303.5088})

\bibitem[\protect\citeauthoryear{{R{\'e}my-Ruyer} et~al.,}{{R{\'e}my-Ruyer}
  et~al.}{2013}]{Remy-Ruyer:2013kx}
{R{\'e}my-Ruyer} A.,  et~al., 2013, \mn@doi [\aap]
  {10.1051/0004-6361/201321602}, \href
  {http://adsabs.harvard.edu/abs/2013A%26A...557A..95R} {557, A95}

\bibitem[\protect\citeauthoryear{{Richards} et~al.,}{{Richards}
  et~al.}{2006}]{Richards:2006fj}
{Richards} G.~T.,  et~al., 2006, \mn@doi [\apjs] {10.1086/506525}, \href
  {http://adsabs.harvard.edu/abs/2006ApJS..166..470R} {166, 470}

\bibitem[\protect\citeauthoryear{{Rosario} et~al.,}{{Rosario}
  et~al.}{2012}]{Rosario:2012fr}
{Rosario} D.~J.,  et~al., 2012, \mn@doi [\aap] {10.1051/0004-6361/201219258},
  \href {http://adsabs.harvard.edu/abs/2012A%26A...545A..45R} {545, A45}

\bibitem[\protect\citeauthoryear{{Roussel}}{{Roussel}}{2013}]{Roussel:2013gf}
{Roussel} H.,  2013, \mn@doi [\pasp] {10.1086/673310}, \href
  {http://adsabs.harvard.edu/abs/2013PASP..125.1126R} {125, 1126}

\bibitem[\protect\citeauthoryear{{Rovilos} et~al.,}{{Rovilos}
  et~al.}{2012}]{Rovilos:2012wd}
{Rovilos} E.,  et~al., 2012, \mn@doi [\aap] {10.1051/0004-6361/201218952},
  \href {http://adsabs.harvard.edu/abs/2012A%26A...546A..58R} {546, A58}

\bibitem[\protect\citeauthoryear{{Rush}, {Malkan}  \& {Edelson}}{{Rush}
  et~al.}{1996}]{Rush:1996db}
{Rush} B.,  {Malkan} M.~A.,   {Edelson} R.~A.,  1996, \mn@doi [\apj]
  {10.1086/178132}, \href {http://adsabs.harvard.edu/abs/1996ApJ...473..130R}
  {473, 130}

\bibitem[\protect\citeauthoryear{{Sanders}, {Soifer}, {Elias}, {Madore},
  {Matthews}, {Neugebauer}  \& {Scoville}}{{Sanders}
  et~al.}{1988}]{Sanders:1988fk}
{Sanders} D.~B.,  {Soifer} B.~T.,  {Elias} J.~H.,  {Madore} B.~F.,  {Matthews}
  K.,  {Neugebauer} G.,   {Scoville} N.~Z.,  1988, \mn@doi [\apj]
  {10.1086/165983}, \href {http://adsabs.harvard.edu/abs/1988ApJ...325...74S}
  {325, 74}

\bibitem[\protect\citeauthoryear{{Scharw{\"a}chter}, {Combes}, {Salom{\'e}},
  {Sun}  \& {Krips}}{{Scharw{\"a}chter} et~al.}{2015}]{Scharwachter:2015ez}
{Scharw{\"a}chter} J.,  {Combes} F.,  {Salom{\'e}} P.,  {Sun} M.,   {Krips} M.,
   2015, preprint, \href {http://adsabs.harvard.edu/abs/2015arXiv150702292S} {}
  (\mn@eprint {arXiv} {1507.02292})

\bibitem[\protect\citeauthoryear{{Scott}}{{Scott}}{1992}]{Scott:1992xy}
{Scott} D.~W.,  1992, {Multivariate Density Estimation}

\bibitem[\protect\citeauthoryear{{Shi} et~al.,}{{Shi}
  et~al.}{2005}]{Shi:2005rc}
{Shi} Y.,  et~al., 2005, \mn@doi [\apj] {10.1086/431344}, \href
  {http://adsabs.harvard.edu/abs/2005ApJ...629...88S} {629, 88}

\bibitem[\protect\citeauthoryear{{Silk} \& {Mamon}}{{Silk} \&
  {Mamon}}{2012}]{Silk:2012fj}
{Silk} J.,  {Mamon} G.~A.,  2012, \mn@doi [Research in Astronomy and
  Astrophysics] {10.1088/1674-4527/12/8/004}, \href
  {http://adsabs.harvard.edu/abs/2012RAA....12..917S} {12, 917}

\bibitem[\protect\citeauthoryear{{Smith} et~al.,}{{Smith}
  et~al.}{2012}]{Smith:2012fj}
{Smith} M.~W.~L.,  et~al., 2012, \mn@doi [\apj] {10.1088/0004-637X/756/1/40},
  \href {http://adsabs.harvard.edu/abs/2012ApJ...756...40S} {756, 40}

\bibitem[\protect\citeauthoryear{{Spinoglio}, {Andreani}  \&
  {Malkan}}{{Spinoglio} et~al.}{2002}]{Spinoglio:2002uq}
{Spinoglio} L.,  {Andreani} P.,   {Malkan} M.~A.,  2002, \mn@doi [\apj]
  {10.1086/340302}, \href {http://adsabs.harvard.edu/abs/2002ApJ...572..105S}
  {572, 105}

\bibitem[\protect\citeauthoryear{{Stern} et~al.,}{{Stern}
  et~al.}{2012}]{Stern:2012mz}
{Stern} D.,  et~al., 2012, \mn@doi [\apj] {10.1088/0004-637X/753/1/30}, \href
  {http://adsabs.harvard.edu/abs/2012ApJ...753...30S} {753, 30}

\bibitem[\protect\citeauthoryear{{Terashima} \& {Wilson}}{{Terashima} \&
  {Wilson}}{2003}]{Terashima:2003fv}
{Terashima} Y.,  {Wilson} A.~S.,  2003, \mn@doi [\apj] {10.1086/345339}, \href
  {http://adsabs.harvard.edu/abs/2003ApJ...583..145T} {583, 145}

\bibitem[\protect\citeauthoryear{{Tombesi}, {Mel{\'e}ndez}, {Veilleux},
  {Reeves}, {Gonz{\'a}lez-Alfonso}  \& {Reynolds}}{{Tombesi}
  et~al.}{2015}]{Tombesi:2015fj}
{Tombesi} F.,  {Mel{\'e}ndez} M.,  {Veilleux} S.,  {Reeves} J.~N.,
  {Gonz{\'a}lez-Alfonso} E.,   {Reynolds} C.~S.,  2015, \mn@doi [\nat]
  {10.1038/nature14261}, \href
  {http://adsabs.harvard.edu/abs/2015Natur.519..436T} {519, 436}

\bibitem[\protect\citeauthoryear{{Tremonti} et~al.,}{{Tremonti}
  et~al.}{2004}]{Tremonti:2004fq}
{Tremonti} C.~A.,  et~al., 2004, \mn@doi [\apj] {10.1086/423264}, \href
  {http://adsabs.harvard.edu/abs/2004ApJ...613..898T} {613, 898}

\bibitem[\protect\citeauthoryear{{Urry} \& {Padovani}}{{Urry} \&
  {Padovani}}{1995}]{Urry:1995il}
{Urry} C.~M.,  {Padovani} P.,  1995, \mn@doi [\pasp] {10.1086/133630}, \href
  {http://adsabs.harvard.edu/abs/1995PASP..107..803U} {107, 803}

\bibitem[\protect\citeauthoryear{{Vasudevan}, {Brandt}, {Mushotzky}, {Winter},
  {Baumgartner}, {Shimizu}, {Schneider}  \& {Nousek}}{{Vasudevan}
  et~al.}{2013}]{Vasudevan:2013dz}
{Vasudevan} R.~V.,  {Brandt} W.~N.,  {Mushotzky} R.~F.,  {Winter} L.~M.,
  {Baumgartner} W.~H.,  {Shimizu} T.~T.,  {Schneider} D.~P.,   {Nousek} J.,
  2013, \mn@doi [\apj] {10.1088/0004-637X/763/2/111}, \href
  {http://adsabs.harvard.edu/abs/2013ApJ...763..111V} {763, 111}

\bibitem[\protect\citeauthoryear{{Veilleux} et~al.,}{{Veilleux}
  et~al.}{2013}]{Veilleux:2013qq}
{Veilleux} S.,  et~al., 2013, \mn@doi [\apj] {10.1088/0004-637X/776/1/27},
  \href {http://adsabs.harvard.edu/abs/2013ApJ...776...27V} {776, 27}

\bibitem[\protect\citeauthoryear{{Weaver} et~al.,}{{Weaver}
  et~al.}{2010}]{Weaver:2010rt}
{Weaver} K.~A.,  et~al., 2010, \mn@doi [\apj] {10.1088/0004-637X/716/2/1151},
  \href {http://adsabs.harvard.edu/abs/2010ApJ...716.1151W} {716, 1151}

\bibitem[\protect\citeauthoryear{{Werner} et~al.,}{{Werner}
  et~al.}{2004}]{Werner:2004cr}
{Werner} M.~W.,  et~al., 2004, \mn@doi [\apjs] {10.1086/422992}, \href
  {http://adsabs.harvard.edu/abs/2004ApJS..154....1W} {154, 1}

\bibitem[\protect\citeauthoryear{{White} et~al.,}{{White}
  et~al.}{2000}]{White:2000rz}
{White} R.~L.,  et~al., 2000, \mn@doi [\apjs] {10.1086/313300}, \href
  {http://adsabs.harvard.edu/abs/2000ApJS..126..133W} {126, 133}

\bibitem[\protect\citeauthoryear{{Winter}, {Mushotzky}, {Reynolds}  \&
  {Tueller}}{{Winter} et~al.}{2009}]{Winter:2009kx}
{Winter} L.~M.,  {Mushotzky} R.~F.,  {Reynolds} C.~S.,   {Tueller} J.,  2009,
  \mn@doi [\apj] {10.1088/0004-637X/690/2/1322}, \href
  {http://adsabs.harvard.edu/abs/2009ApJ...690.1322W} {690, 1322}

\bibitem[\protect\citeauthoryear{{Winter}, {Lewis}, {Koss}, {Veilleux},
  {Keeney}  \& {Mushotzky}}{{Winter} et~al.}{2010}]{Winter:2010yq}
{Winter} L.~M.,  {Lewis} K.~T.,  {Koss} M.,  {Veilleux} S.,  {Keeney} B.,
  {Mushotzky} R.~F.,  2010, \mn@doi [\apj] {10.1088/0004-637X/710/1/503}, \href
  {http://adsabs.harvard.edu/abs/2010ApJ...710..503W} {710, 503}

\bibitem[\protect\citeauthoryear{{Winter}, {Veilleux}, {McKernan}  \&
  {Kallman}}{{Winter} et~al.}{2012}]{Winter:2012yq}
{Winter} L.~M.,  {Veilleux} S.,  {McKernan} B.,   {Kallman} T.~R.,  2012,
  \mn@doi [\apj] {10.1088/0004-637X/745/2/107}, \href
  {http://adsabs.harvard.edu/abs/2012ApJ...745..107W} {745, 107}

\bibitem[\protect\citeauthoryear{{Wright} et~al.,}{{Wright}
  et~al.}{2010}]{Wright:2010fk}
{Wright} E.~L.,  et~al., 2010, \mn@doi [\aj] {10.1088/0004-6256/140/6/1868},
  \href {http://adsabs.harvard.edu/abs/2010AJ....140.1868W} {140, 1868}

\bibitem[\protect\citeauthoryear{{Xu}, {Livio}  \& {Baum}}{{Xu}
  et~al.}{1999}]{Xu:1999ty}
{Xu} C.,  {Livio} M.,   {Baum} S.,  1999, \mn@doi [\aj] {10.1086/301007}, \href
  {http://adsabs.harvard.edu/abs/1999AJ....118.1169X} {118, 1169}

\bibitem[\protect\citeauthoryear{{Xu}, {Rieke}, {Egami}, {Haines}, {Pereira}
  \& {Smith}}{{Xu} et~al.}{2015}]{Xu:2015yq}
{Xu} L.,  {Rieke} G.~H.,  {Egami} E.,  {Haines} C.~P.,  {Pereira} M.~J.,
  {Smith} G.~P.,  2015, \mn@doi [\apj] {10.1088/0004-637X/808/2/159}, \href
  {http://adsabs.harvard.edu/abs/2015ApJ...808..159X} {808, 159}

\bibitem[\protect\citeauthoryear{{di Serego Alighieri} et~al.,}{{di Serego
  Alighieri} et~al.}{2013}]{2013A&A...552A...8D}
{di Serego Alighieri} S.,  et~al., 2013, \mn@doi [\aap]
  {10.1051/0004-6361/201220551}, \href
  {http://adsabs.harvard.edu/abs/2013A%26A...552A...8D} {552, A8}

\makeatother
\end{thebibliography}
}
\end{document}